\documentclass[aop,seceqn,mtplusscr,noautosecdot,dvips]{arximspdf}
\usepackage{mathrsfs}
\usepackage{stmaryrd}

\doi{10.1214/009117906000000872}
\volume{35}
\issue{4}
\pubyear{2007}
\firstpage{1479}
\lastpage{1531}

\makeatletter
\newcommand{\auf}{\llbracket}
\newcommand{\zu}{\rrbracket}
\newcommand{\mal}{\stackrel{\mbox{\tiny$\bullet$}}{}}
\newcommand{\rr}{\mathbb R}
\newcommand{\rp}{\mathbb R _+}
\newcommand{\nn}{\mathbb N}
\newcommand{\Var}{\operatorname{Var}}
\newtheorem{teo}{Theorem}[section]
\newtheorem{prop}[teo]{Proposition}

\newtheorem{lemma}[teo]{Lemma}
\newtheorem{cor}[teo]{Corollary}
\newproclaim{defi}[teo]{Definition}
\newproclaim{exa}[teo]{Example}
\newproclaim{assumption}[teo]{Assumption}

\newproclaim{rem}[teo]{Remark}
\newcommand{\FFF}{\mathscr F}
\newcommand{\SSS}{\mathscr S^2}
\newcommand{\ssl}{\mathscr S^2_{\mathrm{loc}}}
\newcommand{\LLL}{\mathscr L}
\newcommand{\EEE}{\mathscr E}
\newcommand{\PPP}{\mathscr P}
\newcommand{\BBB}{\mathscr B}
\newcommand{\apl}{\mathscr A^+_{\mathrm{loc}}}
\newcommand{\gl}{G_{\mathrm{loc}}}
\newcommand{\lloc}{L^2_{\mathrm{loc}}}
\newcommand{\hloc}{\mathscr H^2_{\mathrm{loc}}}
\makeatother

\begin{document}
\begin{frontmatter}

\title{On the structure of general\break   mean-variance hedging strategies}
\runtitle{Mean-variance hedging}

\begin{aug}
\author[A]{\fnms{Ale\v s} \snm{\v Cern\'y}\ead[label=e1]{cerny@martingales.info}} and
\author[B]{\fnms{Jan} \snm{Kallsen}\corref{}\ead[label=e2]{kallsen@ma.tum.de}}
\runauthor{A. \v Cern\'y and J. Kallsen} \pdfauthor{Ales Cerny, Jan
Kallsen} \affiliation{City University London and Technische
Universit\"at M\"unchen}
\address[A]{Cass Business School\\
City University London\\  106 Bunhill Row\\  London ECIY 8TZ\\ United
Kingdom\\
\printead{e1}} 
\address[B]{HVB-Stiftungsinstitut f\"ur\\
\quad Finanzmathematik\\  Zentrum Mathematik\\  Technische
Universit\"at M\"unchen\\  Boltzmannstra\ss e 3\\  85747 Garching bei
M\"unchen\\  Germany\\  \printead{e2}}
\end{aug}

\received{\smonth{6} \syear{2005}}
\revised{\smonth{10} \syear{2006}}

%
\begin{abstract}
We provide a new characterization of mean-variance hedging strategies
in a general semimartingale market. The key point is the introduction
of a new probability measure $P^\star$ which turns the dynamic asset
allocation problem into a myopic one. The minimal martingale measure
relative to $P^\star$ coincides with the variance-optimal martingale
measure relative to the original probability measure $P$.
\end{abstract}

%

\end{frontmatter}

{\small \begin{center}\textbf{Contents}\end{center}
\begin{enumerate}
\item Introduction
    \begin{enumerate}
    \item[1.1.] Overview
    \item[1.2.] Semimartingale characteristics and notation
    \end{enumerate}

\item Admissible strategies and quadratic hedging
    \begin{enumerate}
    \item[2.1.] Admissible strategies
    \item[2.2.] Mean-variance hedging
    \end{enumerate}

\item On the pure investment problem
    \begin{enumerate}
    \item[3.1.] Opportunity process
    \item[3.2.] Adjustment process
    \item[3.3.] Variance-optimal signed martingale measure
    \item[3.4.] Opportunity-neutral measure
    \item[3.5.] Characterization of $L$ and $\tilde a$
    \item[3.6.] When does $P^\star=P$ hold?
    \item[3.7.] Determination of the opportunity process
    \end{enumerate}

\item  On the pure investment problem
    \begin{enumerate}
    \item[4.1.] Mean value process and pure hedge coefficient
    \item[4.2.] Main results
    \item[4.3.] Connections to the literature
    \end{enumerate}

\item[] Appendix
    \begin{enumerate}
    \item[A.1.] Locally square-integrable semimartingales
    \item[A.2.] $\sigma$-Martingales
    \end{enumerate}

\item[] References\
\end{enumerate}}

\section{\texorpdfstring{Introduction.}{Introduction}}
\subsection{\texorpdfstring{Overview.}{Overview}}
In incomplete market models perfect replication of contingent claims is
typically impossible. A classical way out is to minimize the
\textit{mean squared hedging error}
\[
E \bigl( (v+\vartheta\mal S_T-H )^2 \bigr)
\]
over all reasonable hedging strategies $\vartheta$ and possibly all
initial endowments $v$. Here, the random variable $H$ denotes the
discounted payoff of the claim, the semimartingale $S$ stands for the
discounted price process of the underlying, the dot refers to
stochastic integration, and $T$ is the time horizon. Mathematically
speaking, one seeks to compute the orthogonal projection of $H$ on some
space of stochastic integrals.

This problem has been extensively studied both as far as general theory
as well as concrete results in specific setups are concerned. In order
to render equal justice (or rather injustice) to most contributions, we
refer the reader to \cite{pham00} and \cite{schweizer99} for
excellent overviews of the literature. More recent publications in this
context include \cite{arai04,arai05,benthal03,%
bobrovnytskaschweizer04,cerny04b,cerny05,cernykallsen06bwp,cernykallsen06awp,%
dinunno02,hipptaksar05,hobson04,houkaratzas04,hubalekal05,lim04,lim05,%
maniatevzadze00,maniatevzadze03b,maniatevzadze03a,sekine05}.

The purpose of this piece of research is to provide a deeper
understanding of the structure of the mean-variance hedging problem in
a general semimartingale context. More specifically, we aim at concrete
formulas for the objects of interest---to the extent that this is
possible without restricting to more specific situations.

If $S$ is a square-integrable martingale, the answer to the above
hedging problem is provided by the \textit{Galtchouk--Kunita--Watanabe
decomposition} of the claim (cf.~\cite{foellmersondermann86}). In
particular, the optimal hedge $\vartheta$ is of the form
\begin{equation}\label{e:kunitawatanabe}
\vartheta_t={d\langle V,S\rangle
_t\over
d\langle S,S\rangle_t},
\end{equation}
where $V_t=E(H|\FFF_t)$ denotes the martingale generated by the
contingent claim~$H$.

If $S$ fails to be a martingale, the hedging problem becomes much more
involved. Relatively explicit results have been obtained by
Schweizer~\cite{schweizer94} under the condition of
\textit{deterministic mean-variance tradeoff}, which can be intepreted
as a certain homogeneity property of the asset price process $S$. In
this case the optimal hedge is the sum of two terms. The first
satisfies an equation resembling (\ref{e:kunitawatanabe}). The second
can be interpreted in terms of a pure investment problem under
quadratic utility.

In the current paper we reduce the general case to the expressions of
\cite{schweizer94}. This is done by a specific nonmartingale change of
measure. If the formulas of \cite{schweizer94} are evaluated relative
to the new \textit{opportunity-neutral measure} $P^\star$ rather than
$P$, they yield the optimal hedge relative to the original probability
measure $P$. We discuss the links to the literature more thoroughly in
Section \ref{su:link}.

The paper is structured as follows. Section \ref{s:admissible} explains
the setup of the mean-variance problem at hand. In particular, we
define a notion of admissibility which ensures the existence of an
optimal hedge. The measure change alluded to above and related objects
are introduced in Section \ref{s:pure}. Subsequently, we turn to the
hedging problem itself. Finally, the appendix contains and summarizes
auxiliary statements on semimartingales. In particular, we prove a
sufficient condition for square integrability of exponential
semimartingales which is needed in Section \ref{s:hedge}.

\subsection{\texorpdfstring{Semimartingale characteristics and
notation.}{Semimartingale characteristics and notation}} Unexplained
notation is typically used as in \cite{js87}. Superscripts refer
generally to coordinates of a vector or vector-valued process rather
than powers. The few exceptions should be obvious from the context. If
$X$ is a semimartingale, $L(X)$ denotes the set of $X$-integrable
predictable processes in the sense of \cite{js87}, III.6.17.

In the subsequent sections, optimal hedging strategies are expressed in
terms of semimartingale characteristics.

\begin{defi}\label{d:bcfa}
Let $X$ be an $\rr^d$-valued semimartingale with characteristics
$(B,C,\nu)$ relative to some truncation function $h:\rr^d\to\rr^d$. By
\cite{js87}, II.2.9 there exists some predictable process $A\in\apl$,
some predictable $\rr^{d\times d}$-valued process $c$ whose values are
nonnegative, symmetric matrices, and some transition kernel $F$ from
$(\Omega\times\rp,\PPP)$ into $(\rr^d,\BBB^d)$ such that
\begin{eqnarray}\label{e:bcfa}
\nonumber B_t=b\mal A_t,\qquad   C_t=c\mal A_t,\qquad   \nu([0,t]\times
G)=F(G)\mal A_t
\\
\eqntext{\mbox{for }t\in[0,T], G\in\BBB^d.}
\end{eqnarray}
We call $(b,c,F,A)$ \textit{differential characteristics} of $X$.
\end{defi}

One should observe that the differential characteristics are not
unique: for example, $(2b,2c,2F,{1\over2}A)$ yields another version. Especially
for $A_t=t$, one can interpret $b_t$ or rather $b_t+\int
(x-h(x))F_t(dx)$ as a drift rate, $c_t$ as a diffusion coefficient, and
$F_t$ as a local jump measure. The differential characteristics are
typically derived from other ``local'' representations of the process,
for example, in terms of a stochastic differential equation.

From now on, we choose the same fixed process $A$ for all the (finitely
many) semimartingales in this paper. The results do not depend on its
particular choice. In concrete models, $A$ is often taken to be $A_t=t$
(e.g., for L\'evy processes, diffusions, It\^o processes, etc.) and
$A_t=[t]:=\max\{n\in\nn\dvtx n\leq t\}$ (discrete-time processes). Since
almost all semimartingales of interest in this paper are actually
special semimartingales, we use from now on the (otherwise forbidden)
``truncation'' function
\[
h(x):=x,
\]
which simplifies a number of expressions considerably.

By $\langle X,Y\rangle$ we denote the $P$-compensator of $[X,Y]$
provided that $X,Y$ are semimartingales such that $[X,Y]$ is
$P$-special (cf. \cite{jacod79}, page~37). If $X$ and $Y$ are
vector-valued, then $[X,Y]$ and $\langle X,Y\rangle$ are to be
understood as matrix-valued processes with components $[X^i,Y^j]$ and
$\langle X^i,Y^j\rangle$, respectively. Moreover, if both $Y$ and a
predictable process $\vartheta$ are $\rr ^d$-valued, then the notation
$\vartheta\mal[X,Y]$ (and accordingly $\vartheta\mal\langle
X,Y\rangle$) refers to the vector-valued process whose components
$\vartheta\mal[X^i,Y]$ are the vector-stochastic integral of
$(\vartheta^j)_{j=1,\dots,d}$ relative to $([X^i,Y^j])_{j=1,\dots,d}$.
If $P^\star$ denotes another probability measure, we write $\langle
X,Y\rangle^{P^\star}$ for the $P^\star$-compensator of $[X,Y]$.

In the whole paper, we write $M^X$ for the local martingale part and
$A^X$ for the predictable part of finite variation in the canonical
decomposition
\[
X=X_0+M^X+A^X
\]
of a special semimartingale $X$.
If $P^\star$ denotes another probability measure, we write accordingly
\[
X=X_0+M^{X\star}+A^{X\star}
\]
for the $P^\star$-canonical decomposition of $X$.

If $(b,c,F,A)$ denote differential characteristics of an $\rr^d$-valued
special semimartingale $X$, we use the notation $\tilde c, \hat c$ for
\textit{modified second characteristics} in the following sense
(provided that the integrals exist):
\begin{eqnarray}
\tilde c &:=& c+\int xx^\top F(dx),\label{e:mod}
\\
\hat c&:=&c+\int xx^\top F(dx)-bb^\top\Delta A.\label{e:modi}
\end{eqnarray}
Observe that $x^\top\hat cx\leq x^\top\tilde cx$ for any $x\in\rr^d$.
The notion of modified second characteristics is motivated by the
following:

\begin{prop}\label{p:spitzklammer}
Let $X$ be an $\rr^d$-valued special semimartingale with differential
characteristics $(b,c,F,A)$ and modified second characteristics as in
\textup{(\ref{e:mod})} and \textup{(\ref{e:modi})}. If the
corresponding integrals exist, then
\begin{eqnarray*}
\langle X,X\rangle&=&\tilde c\mal A,
\\
\langle{M^X},{M^X}\rangle&=&\hat c\mal A.
\end{eqnarray*}
\end{prop}

\begin{pf}
The first equation follows from \cite{js87}, I.4.52, the second from
\cite{js87}, II.2.17 (adjusted for the truncation function).
\end{pf}

From now on we use the notation $(b^X,c^X,F^X,A)$ to denote
differential characteristics of a special semimartingale $X$.
Accordingly, $\tilde c^X$, $\hat c^X$ stands for the modified second
characteristics of $X$. If they refer to some probability measure
$P^\star$ rather than $P$, we write instead $(b^{X\star
},c^{X\star},F^{X\star},A)$ and $\tilde c^{X\star}, \hat c^{X\star}$,
respectively. We denote the joint characteristics of two special
semimartingales $X,Y$ [i.e., the characteristics of $(X,Y)$] as
\[
(b^{X,Y},c^{X,Y},F^{X,Y},A )= \left(\pmatrix{b^X\cr b^Y}, \pmatrix{
c^{X} & c^{XY} \cr   c^{YX} & c^{Y}}, F^{X,Y},A \right)
\]
and
\[
\tilde c^{X,Y}= \pmatrix{ \tilde c^{X} & \tilde c^{XY} \cr   \tilde
c^{YX} & \tilde c^{Y}},\qquad \hat c^{X,Y}= \pmatrix{ \hat c^{X} & \hat
c^{XY} \cr   \hat c^{YX} & \hat c^{Y}} .
\]

In the whole paper, we write $c^{-1}$ for the Moore--Penrose
pseudoinverse of a matrix or matrix-valued process $c$, which is a
particular matrix satisfying $cc^{-1}c=c$ (cf. \cite{albert72}). From
the construction it follows that the mapping $c\mapsto c^{-1}$ is
measurable. Moreover, $c^{-1}$ is nonnegative and symmetric if this
holds for $c$.

Finally, we write $X\sim Y$ (resp. $X\sim^\star Y$) if two
semimartingales differ only by some $P$-$\sigma$-martingale (or some
$P^\star$-$\sigma$-martingale, resp.). Some facts on
\mbox{$\sigma$-}martingales are summarized in Appendix \ref{su:sigma}.

\section{\texorpdfstring{Admissible strategies and quadratic hedging.}{Admissible strategies and quadratic hedging}} \label{s:admissible}
We work on a filtered probability space $(\Omega,\FFF,(\FFF_t)_{t\in
[0,T]},P)$, where $T\in\rp$ denotes a fixed terminal time. The
$\rr^d$-valued process $S=(S^1_t,\dots ,S^d_t)_{t\in[0,T]}$ represents
the discounted prices of $d$ securities. We assume that
\begin{equation}\label{e:L2bounded}
\sup \{E ((S^i_\tau)^2 )\dvtx \tau\mbox{ stopping time }, i=1,\dots,d
\}<\infty,
\end{equation}
that is, $S$ is a \textit{$L^2(P)$-semimartingale} in the sense of
\cite{delbaenschachermayer96}.

Moreover, we make the following standing:

\begin{assumption} \label{a:eins}
There exists some equivalent $\sigma$-martingale measure with
square-integrable density, that is, some probability measure $Q\sim P$
with $E(({dQ\over dP})^2)<\infty$ and such that $S$ is a
$Q$-$\sigma$-martingale.
\end{assumption}

This can be interpreted as a natural no-free-lunch condition in the
present quadratic context. More specifically, Th\'eor\`eme 2 in
\cite{stricker90} and standard arguments show that Assumption
\ref{a:eins} is equivalent to the \textit{absence of $L^2$-free
lunches} in the sense that
\[
\overline{K^s_2(0)-L^2_+}\cap L^2_+=\{0\},
\]
where $K^s_2(0)$ denotes the set of payoffs of simple trading defined
below, $L^2_+$ contains the nonnegative square-integrable random
variables, and the closure is to be taken in $L^2(P)$.

\subsection{\texorpdfstring{Admissible strategies.}{Admissible strategies}}\label{su:admissible}
The choice of the set of admissible trading strategies in continuous
time is a delicate point. If it is too large, arbitrage opportunities
occur even in the Black--Scholes model, if it is too small, optimal
strategies as, for example, the replicating portfolio of a European
call in the Black--Scholes model fail to exist. Inspired by Delbaen and
Schachermayer \cite{delbaenschachermayer96}, we consider the closure
(in a proper $L^2$-sense) of the set of simple strategies.

More specifically, an $\rr^d$-valued process $\vartheta$ is called
\textit{simple} if it is a linear combination of processes of the form
$Y1_{\zu\tau_1,\tau_2\zu}$, where $\tau_1\leq\tau_2$ denote stopping
times and $Y$ a bounded $\FFF_{\tau_1}$-measurable random variable. We
call a payoff \textit{attainable by simple trading with initial
endowment $v\in L^2(\Omega,\FFF_0,P)$} if it belongs to the set
\[
K^s_2(v):=\{v+\vartheta\mal S_T\dvtx \vartheta\mbox{ simple}\}.
\]
If the initial endowment $v$ is not fixed beforehand,
we consider instead the set
\[
K^s_2(\FFF_0):=\{v+\vartheta\mal S_T\dvtx v\in
L^2(\Omega,\FFF_0,P),\vartheta \mbox{ simple}\}.
\]

Since the hedging problems in this paper concern the approximation of
arbitrary payoffs $H$ in $L^2(P)$ by attainable outcomes, it makes
perfect sense from an economical point of view to call the elements of
the $L^2(P)$-closures $K_2(v):=\overline{K^s_2(v)}$, respectively,
$K_2(\FFF_0):=\overline{K^s_2(\FFF_0)}$ \textit{attainable} as well.
These outcomes can be written as a stochastic integral $v+\vartheta\mal
S_T$ with some strategy $\vartheta\in L(S)$ that can be approximated in
the following sense by simple strategies (cf. Lemmas~\ref{l:k20} and
\ref{l:k2f0} below).

\begin{defi}\label{d:admissible}
We call $\vartheta\in L(S)$ \textit{admissible strategy} if there
exists some sequence $(\vartheta^{(n)})_{n\in\nn}$ of simple strategies
such that
\begin{eqnarray*}
\vartheta^{(n)}\mal S_t&\to& \vartheta\mal S_t\qquad\mbox{in probability for
any }t\in[0,T]\quad\mbox{and}
\\
\vartheta^{(n)}\mal S_T&\to&
\vartheta\mal S_T\qquad\mbox{in }L^2(P).
\end{eqnarray*}
Similarly, we call $(v,\vartheta)\in L^0(\Omega,\FFF_0,P)\times L(S)$
\textit{admissible endowment/strategy pair} if there exist some
sequences $(v^{(n)})_{n\in\nn}$ in $L^2(\Omega,\FFF_0,P)$ and
$(\vartheta^{(n)})_{n\in\nn}$ of simple strategies such that
\begin{eqnarray*}
v^{(n)}+\vartheta^{(n)}\mal S_t&\to& v+\vartheta\mal S_t\qquad\mbox{in
probability for any }t\in[0,T]\quad\mbox{and}
\\
v^{(n)}+\vartheta^{(n)}\mal S_T&\to& v+\vartheta\mal S_T\qquad\mbox{in
}L^2(P).
\end{eqnarray*}
We set
\begin{eqnarray*}
\overline\Theta&:=&\{\vartheta\in L(S)\dvtx  \vartheta\mbox{ admissible}\},
\\
\overline{L^2(\FFF_0)\times\Theta} &:=&\{(v,\vartheta)\in
L^0(\Omega,\FFF_0,P)\times L(S)\dvtx  (v,\vartheta) \mbox{
admissible}\}.
\end{eqnarray*}
\end{defi}

One easily verifies that
$\overline{L^2(\FFF_0)\times\Theta}=\rr\times\overline\Theta$ if the
initial $\sigma$-field $\FFF_0$ is trivial. Admissible strategies are
linked via duality to martingale measures of the following kind:

\begin{defi}\label{d:SSMM}
We call a signed measure $Q\ll P$ with $Q(\Omega)=1$ \textit{absolutely
continuous signed $\sigma$-martingale measure (S$\sigma$MM)} if $SZ^Q$
is a \mbox{$P$-$\sigma$-}mart\-ingale for the density process
\[
\label{e:densityprocess} Z^Q_t:=E \biggl( {dQ\over dP} \bigg|\FFF _t
\biggr)
\]
of~$Q$.
\end{defi}

A probability measure $Q\sim P$ is a S$\sigma$MM if and only if $S$ is
a $Q$-$\sigma$-martingale (cf. Lemma~\ref{l:Pstarmartingale}).

\begin{lemma}\label{l:k20}
For $H\in L^2(P)$ and $v\in L^2(\Omega,\FFF_0,P)$ the following
statements are equivalent:
\begin{enumerate}
\item $H\in K_2(v)$.

\item $E_Q(H-v)=0$ for any S$\sigma$MM $Q$ with ${dQ\over dP}\in
L^2(P)$.

\item $H=v+\vartheta\mal S_T$ with some $\vartheta\in\overline\Theta$.

\item $H=v+\vartheta\mal S_T$ with some $\vartheta\in L(S)$ such that
$(\vartheta\mal S)Z^Q$ is a martingale for any S$\sigma$MM $Q$ with
density process $Z^Q$ and ${dQ\over dP}\in L^2(P)$.
\end{enumerate}
In particular, we have $K_2(v)=\{v+\vartheta\mal S_T\dvtx
\vartheta\in\overline \Theta\}$.
\end{lemma}

\begin{pf}
It suffices to consider the case $v=0$.

1${}\Rightarrow{}$3, 4: \textit{Step} 1: We start by showing that
statement 4 holds for $H\in K_2^s(0)$, that is, for $H=\vartheta\mal
S_T$ with some simple $\vartheta$. Integration by parts yields
\begin{eqnarray}\label{e:partial}
(\vartheta\mal S)Z^Q &=&(\vartheta\mal S)_-\mal
Z^Q+\vartheta\mal(Z^Q_-\mal S+[Z^Q,S])
\nonumber
\\[-8pt]
\\[-8pt]
\nonumber
&=&(\vartheta\mal S_--\vartheta^\top S_-)\mal Z^Q+\vartheta\mal(SZ^Q),
\end{eqnarray}
which implies that $(\vartheta\mal S)Z^Q$ is a $\sigma$-martingale.
Since $\sup_{t\in[0,T]}|Z^Q_t|\in L^2(P)$ by Doob's inequality and
$\vartheta\mal S$ is a $L^2$-semimartingale in the sense of
(\ref{e:L2bounded}), we have that $(\vartheta\mal S)Z^Q$ is of class
(D) and hence a martingale (cf.  Lemma~\ref{l:martingale}).

\textit{Step} 2: Let $H^n=\vartheta^{(n)}\mal S_T$ be an approximating
sequence for $H\in K_2(0)$. From~\cite{delbaenschachermayer96},
Theorem~1.2, it follows that $H$ has a representation $H=\vartheta\mal
S_T$ for some $\vartheta\in L(S)$. In the proof of this theorem it is
actually shown that $\vartheta$ can be chosen such that
$\vartheta^{(n)}\mal S_t$ converges in probability to $\vartheta\mal
S_t$ for any $t\in[0,T]$.

Since $H^nZ^Q_T\to HZ^Q_T$ in $L^1(P)$, we have that
\[
E\bigl(\bigl(\vartheta^{(n)}\mal S_T\bigr)Z^Q_T|\FFF_t\bigr)\to
E\bigl((\vartheta\mal S_T)Z^Q_T|\FFF_t\bigr)
\]
in $L^1(P)$ and hence in probability. Step 1 yields
$E((\vartheta^{(n)}\mal S_T)Z^Q_T|\FFF_t)=(\vartheta^{(n)}\mal
S_t)Z^Q_t.$ Together, it follows that $E((\vartheta\mal
S_T)Z^Q_T|\FFF_t)=(\vartheta\mal S_t)Z^Q_t$.

3${}\Rightarrow{}$1: This is obvious.

4${}\Rightarrow{}$2: This is obvious as well.

2${}\Rightarrow{}$1: It suffices to show that $K_2(0)^\perp\subset
(V^\perp )^\perp$ for
\[
V:= \biggl\{{dQ\over dP}\dvtx Q\mbox{ S$\sigma$MM with }{dQ\over dP}\in
L^2(P) \biggr\},
\]
where the orthogonal complements refer to $L^2(P)$. Let $Y\in
K_2(0)^\perp$ and set $Z_t:= E(Y|\FFF_t)$. For $s\leq t$ and
$F\in\FFF_s$ we have
\begin{eqnarray*}
&& E\bigl(1_F(S_tZ_t-S_sZ_s)\bigr)
\\
&&\qquad = E\bigl(1_F(S_t-S_s)Y\bigr)-E\bigl(1_FS_t(Z_T-Z_t)\bigr)+E\bigl(1_FS_s(Z_T-Z_s)\bigr)
\\
&&\qquad = 0
\end{eqnarray*}
because $Z$ is a martingale and $1_{F\times(s,t]}\mal S_T\in K_2(0)$.
If $E(Y)\not=0$, then $Y$ is a multiple of a S$\sigma$MM and hence in
$(V^\perp)^\perp$. If $E(Y)=0$, then $Y+{dQ\over dP}\in
V\subset(V^\perp)^\perp$ for the S$\sigma$MM $Q$ from Assumption
\ref{a:eins}, which implies that $Y\in(V^\perp )^\perp$ as
well.
\end{pf}

This leads to the following characterization of admissible strategies:

\begin{cor}\label{co:theta}
We have equivalence between\textup{:}
\begin{enumerate}
\item $\vartheta$ is an admissible strategy.

\item $\vartheta\in L(S)$, $\vartheta\mal S_T\in L^2(P)$, and
$(\vartheta\mal S)Z^Q$ is a martingale for any S$\sigma$MM $Q$ with
density process $Z^Q$ and ${dQ\over dP}\in L^2(P)$.
\end{enumerate}
\end{cor}

\begin{pf}
1${}\Rightarrow{}$2: This follows from the argument in step 2 of the
proof of Lemma~\ref{l:k20}.

2${}\Rightarrow{}$1: We have $\vartheta\mal S_T\in K_2(0)$ by
Lemma~\ref{l:k20}. Let $Q$ be a $\sigma$-martingale measure as in
Assumption \ref{a:eins}. By the proof of Lemma~\ref{l:k20}
(1${}\Rightarrow{}$3,4) there exists some $\tilde\vartheta\in\overline
\Theta$ such that $\tilde\vartheta\mal S$ is a $Q$-martingale with
$\tilde\vartheta \mal S_T=\vartheta \mal S_T$. Since $\vartheta\mal S$
is a $Q$-martingale as well, we have $\tilde \vartheta \mal
S=\vartheta\mal S$ and hence $\vartheta\in\overline\Theta$.
\end{pf}

In the case without fixed initial endowment we have:

\begin{lemma}\label{l:k2f0}
There exists
\begin{enumerate}
\item $K_2(\FFF_0)= \{v+\vartheta\mal S_T\dvtx
(v,\vartheta)\in\overline{L^2(\FFF _0)\times \Theta} \}$.

\item If $(v,\vartheta)\in\overline{L^2(\FFF_0)\times\Theta}$, then
$(v+\vartheta\mal S)Z^Q$ is a martingale for any S$\sigma$MM $Q$ with
density process $Z^Q$ and ${dQ\over dP}\in L^2(P)$.
\end{enumerate}
\end{lemma}

\begin{pf}
This follows by rather obvious extension of the proof of
Lemma~\ref{l:k20} (1${}\Rightarrow{}$3, 4) and the underlying arguments
in \cite{delbaenschachermayer96}.
\end{pf}

\begin{rem}\label{r:erweiterung}
An inspection of the proof reveals that statement 2 in
Corollary~\ref{co:theta} and Lemma~\ref{l:k2f0} holds for any
square-integrable martingale $Z^Q$ such that $SZ^Q$ is a
$\sigma$-martingale, that is, the property $E(Z^Q_T)=1$ is not needed.
\end{rem}

If necessary the whole setup can be relaxed to slightly more general
price processes:

\begin{rem}\label{r:general}
Instead of (\ref{e:L2bounded}), Delbaen and Schachermayer
\cite{delbaenschachermayer96} assume only that $S$ is a \textit{local
\mbox{$L^2(P)$-}semimartingale}, that is, that there is a localizing
sequence of stopping times $(U_n)_{n\in\nn}$ such that:
\[
\sup \{E ((S^i_\tau)^2 )\dvtx \tau\leq U_n \mbox{ stopping time, }
i=1,\dots,d \}<\infty
\]
for any $n\in\nn$. Equivalently, $S^1,\dots,S^d$ are locally
square-integrable semimartingales (cf. Definition~\ref{d:s2} and
Lemma~\ref{l:s2} in the \hyperref[app]{Appendix}). In this case Delbaen
and Schachermayer \cite{delbaenschachermayer96} call a linear
combination of strategies $Y1_{\zu \tau _1,\tau_2\zu}$ \textit{simple}
if the corresponding stopping times $\tau_1\leq\tau_2$ are dominated by
some $U_n$. One easily verifies that \textit{all results in this paper
extend to this slightly more general setup}.

The corresponding admissible sets $\overline\Theta$ and $\overline
{L^2(\FFF _0)\times \Theta}$ from Definition~\ref{d:admissible} do not
depend on the chosen sequence $(U_n)_{n\in\nn}$: For $\overline\Theta$
this follows from the characterization in Corollary \ref{co:theta}.
Moreover, $K_2(\FFF_0)=\overline{L^2(\Omega,\FFF_0,P)+K_2(0)}$ does not
depend on $(U_n)_{n\in\nn}$ by Lemma~\ref{l:k20}. Using
Lemma~\ref{l:k2f0} and arguing similarly as in the proof of Corollary
\ref{co:theta} (2${}\Rightarrow{}$1), we have that the same is true for
$\overline{L^2(\FFF_0)\times\Theta}$.
\end{rem}

Many results in the subsequent sections could also be expressed in
terms of the generally different set of strategies considered in
\cite{schweizer94} and other papers on mean-variance hedging, namely
\[
\Theta:= \{\vartheta\in L(S)\dvtx \vartheta\mal S\in\SSS \},
\]
where $\SSS$ denotes the set of square-integrable semimartingales (cf.
Definition~\ref{d:s2}). In contrast to $\{v+\vartheta\mal S_T\dvtx
\vartheta\in\overline\Theta\}$, the set $\{v+\vartheta\mal S_T\dvtx
\vartheta\in\Theta\}$ is not necessarily closed. This issue is
discussed in detail by Monat and Stricker~\cite{monatstricker95},
Delbaen et al. \cite{delbaenal97} and Choulli, Krawczyk and Stricker
\cite{choullial98}. By considering \mbox{$L^2$-}closures in the above
sense, one avoids the problem that optimal hedging strategies may fail
to exist. In the context of continuous processes, our notion of
admissible strategies coincides with the one of Gourieroux, Laurent and
Pham \cite{gourierouxal98} and  Laurent and Pham \cite{laurentpham99}.
Recently, the question of how to choose a reasonable set of strategies
in a quadratic context has been discussed by Xia and Yan
\cite{xiayan06}. Their notion of admissibility differs from ours but
their set of terminal payoffs coincides with $K_2(0)$.

The relationship between $\overline\Theta$ and $\Theta$ is clarified by
the following result. The first assertion is inspired by a similar
statement in Grandits and Rheinl\"aender \cite{granditsrheinlaender02},
Lemma~2.1 for continuous processes. Loosely speaking, it says that
$\overline\Theta$ is a kind of $L^2$-closure of $\Theta$.

\begin{cor}\label{co:theta2}
We have
\begin{enumerate}
\item $\Theta\subset\overline\Theta$ and $\overline{\{\vartheta\mal
S_T\dvtx \vartheta\in\Theta\}}=K_2(0)=\{\vartheta \mal S_T\dvtx
\vartheta \in\overline\Theta\}$.

\item ${L^2(\Omega,\FFF_0,P)\times\Theta}\subset\overline{L^2(\FFF
_0)\times\Theta }$ and
\begin{eqnarray*}
&& \overline{\{v+\vartheta\mal S_T\dvtx v\in
L^2(\Omega,\FFF_0,P),\vartheta\in \Theta \} }
\\
&&\qquad =K_2(\FFF_0) =\{v+\vartheta\mal S_T\dvtx
(v,\vartheta)\in\overline{L^2(\FFF_0)\times\Theta }\}.
\end{eqnarray*}
\end{enumerate}
In both cases the closure $\overline{\{\cdots\}}$ refers to the
$L^2(P)$-norm.
\end{cor}

\begin{pf}%
1. For $\vartheta\in\Theta$ we have $E(\sup_{t\in[0,T]}|\vartheta\mal
S_t|^2)<\infty$ by Protter \cite{protter04}, Theorem~IV.5.
$\vartheta\in\overline\Theta$ now follows easily from Corollary
\ref{co:theta} (2${}\Rightarrow{}$1) together with (\ref{e:partial})
and Lemma~\ref{l:martingale}. The second equality is shown in
Lemma~\ref{l:k20}. In order to verify the first equality, it suffices
to prove that any simple strategy is in $\Theta$. This may not be true
in the first place. But if the sequence $(U_n)_{n\in\nn}$ in
Remark~\ref{r:general} is chosen such that $(S^i)^{U_n}\in\SSS$ for
$n\in\nn$, $i=1,\dots,d$, then $\vartheta \in\Theta$ for any
simple~$\vartheta$. Since $\overline\Theta$ does not depend on the
chosen sequence $(U_n)_{n\in\nn}$, the claim follows.

2. By statement 1 we have
\[
{L^2(\Omega,\FFF_0,P)\times\Theta}\subset L^2(\Omega,\FFF_0,P)\times
\overline \Theta \subset\overline{L^2(\FFF_0)\times\Theta}.
\]
The equalities follow similarly as in statement 1, this time using
Lemma~~\ref{l:k2f0}.
\end{pf}

\subsection{\texorpdfstring{Mean-variance hedging.}{Mean-variance
hedging}} The goal of this paper is to hedge a fixed contingent claim
with discounted payoff $H\in L^2(\Omega,\FFF,P)$. We consider two
closely related optimization problems.

\begin{defi}\label{d:hedge}
1. We call an admissible endowment/strategy pair $(v_0,\varphi)$
\textit{optimal} if $(v,\vartheta)=(v_0,\varphi)$ minimizes the
expected squared hedging error
\begin{equation}\label{e:error}
E \bigl( (v+\vartheta\mal S_T-H )^2 \bigr)
\end{equation}
over all admissible endowment/strategy pairs $(v,\vartheta)$.

2. If the initial endowment $v=v_0\in L^2(\Omega,\FFF_0,P)$ is given
beforehand, a minimizer $\vartheta=\varphi$ of (\ref{e:error}) over all
$\vartheta\in\overline \Theta$ is called \textit{optimal hedging
strategy for given initial endowment $v_0$}.
\end{defi}
Due to the chosen notion of admissibility, optimal hedges always exist:

\begin{lemma}\label{l:unique}
There exist optimal hedges in the sense of
Definition~\textup{\ref{d:hedge}(1)} and \textup{(2)}. In both cases,
the value process $v_0+\varphi\mal S$ of the optimal hedge is unique up
to a $P$-null set.
\end{lemma}

\begin{pf}The existence follows from Lemmas~\ref{l:k20},
\ref{l:k2f0} and the closedness of $K_2(\FFF_0)$ and $K_2(v_0)$,
respectively.

Denote by $v_0+\varphi\mal S$ and $\tilde v_0+\tilde\varphi\mal S$
value processes of two optimal hedges [which implies that $v_0=\tilde
v_0$ in the situation of Definition~\ref{d:hedge}(2)]. A simple
convexity argument yields $v_0+\varphi\mal S_T=\tilde v_0+\tilde
\varphi \mal S_T$.  It remains to be shown that this implies
$v_0+\varphi\mal S=\tilde v_0+\tilde\varphi\mal S$ up to a $P$-null
set. Otherwise, there exists some $n\in\nn$ such that $P(\tau<T)>0$ for
the stopping time
\[
\tau:=\inf \biggl\{t\in[0,T]\dvtx v_0+\varphi\mal S_t\geq\tilde
v_0+\tilde\varphi \mal S_t+{1\over n} \biggr\}\wedge T
\]
(or possibly with exchanged roles of $\varphi,\tilde\varphi$). From
Corollary \ref{co:theta} and Lemma~\ref{l:k2f0} if follows that
$M:=v_0-\tilde v_0+(\varphi-\tilde\varphi)\mal S$ is a martingale with
respect to the $\sigma$-martingale measure $Q$ from Assumption
\ref{a:eins}. Consequently, $E_Q(M_\tau)=E_Q(M_T)=0$, which is
impossible if $P(\tau<T)>0$.
\end{pf}

\section{\texorpdfstring{On the pure investment problem.}{On the pure investment problem}}\label{s:pure}
In many papers the mean-variance hedging problem is partially reduced
to pure portfolio optimization with quadratic utility. This is done
here as well.

\subsection{\texorpdfstring{Opportunity process.}{Opportunity process}}
In the spirit of Markowitz, we call an admissible strategy $\lambda
^{(\tau)}$ \textit{efficient} on a stochastic interval $\zu\tau,T\zu$
if it minimizes
\begin{equation}\label{e:efficient}
E \bigl((1-\vartheta\mal S_T)^2 \bigr)
\end{equation}
over all $\vartheta\in\overline\Theta$ vanishing on $\auf0,\tau\zu$.
Indeed, by standard arguments there exists no strategy with at most the
same variance yielding a higher expected return. Alternatively, one may
view $\lambda^{(\tau)}$ as optimal hedging strategy on $\zu\tau,T\zu$
for the constant option $H=1$. A crucial role will be played by the
related \textit{opportunity process}
\[
L_t=E \bigl( \bigl(1-\lambda^{(t)}\mal S_T\bigr)^2 |\FFF_t \bigr),
\]
whose existence and properties are yet to be derived.
\begin{lemma}\label{l:existenz}
\textup{1.} For any stopping time $\tau$ there exists an efficient
strategy $\lambda^{(\tau)}$ on $\zu\tau ,T\zu$. Its value process
$1-\lambda^{(\tau)}\mal S$ is uniquely determined.
{\smallskipamount=0pt
\begin{longlist}[2.]
\item[2.] $1-\lambda^{(\varrho)}\mal S_\tau=(1-\lambda^{(\varrho)}\mal
S_\sigma) (1-\lambda^{(\sigma)}\mal S_\tau)$ for all stopping times
$\varrho\leq\sigma\leq\tau$.

\item[3.] If $1-\lambda^{(\sigma)}\mal S_\tau=0$, then $1-\lambda
^{(\sigma )}\mal S_T=0$ for all stopping times $\sigma\leq\tau$.

\item[4.] $E((1-\lambda^{(\tau)}\mal S_T)^2|\FFF_\sigma)\leq
E((1-\vartheta \mal S_T)^2|\FFF_\sigma)$ for all stopping times
$\sigma\leq\tau$ and any strategy $\vartheta\in\overline\Theta$ with
$\vartheta1_{\auf0,\tau\zu}=0$.

\item[5.] $E(1-\lambda^{(\tau)}\mal
S_T|\FFF_\sigma)=E((1-\lambda^{(\tau )}\mal S_T)^2|\FFF_\sigma)
\in(0,1]$ almost surely for all stopping times
$\sigma\leq\tau$.
\end{longlist}}
\end{lemma}

\begin{pf}
1. If $G$ denotes the orthogonal projection of $1$ on
\[
\overline{\{\vartheta\mal S_T\dvtx \vartheta\in\overline\Theta \mbox{
and }\vartheta1_{\auf0,\tau\zu}=0\}}\subset K_2(0),
\]
then there is a sequence $(\vartheta^{(n)})_{n\in\nn}$ of strategies in
$\overline\Theta$ that vanish on $\auf0,\tau\zu$ and satisfy
$\vartheta^{(n)}\mal S_T\to G$ in $L^2(P)$. By Lemma~\ref{l:k20} we
have $G=\vartheta\mal S_T$ for some $\vartheta\in \overline \Theta$.
Moreover, $\vartheta^{(n)}\mal S_T\to\vartheta\mal S_T$ in $L^1(Q)$ for
the $\sigma$-martingale measure $Q$ from Assumption \ref{a:eins}. This
implies $0=\vartheta^{(n)}\mal S_t\to\vartheta\mal S_t$ in $L^1(Q)$
because both $\vartheta^{(n)}\mal S$ and $\vartheta\mal S_t$ are
$Q$-martingales by Corollary \ref{co:theta}. Hence we have
$\vartheta1_{\auf0,\tau\zu}=0$ without loss of generality. Uniqueness
follows as in the proof of Lemma~\ref{l:unique}. {\smallskipamount=0pt
\begin{longlist}[2.]
\item[2.] We start by showing that
\begin{eqnarray}\label{e:kleiner}
&& E \bigl( \bigl(1-\lambda^{(\varrho )}\mal S_T\bigr)^2 |\FFF_\sigma
\bigr) \nonumber
\\[-9.5pt]
\\[-9.5pt]
\nonumber &&\qquad \leq E \bigl( \bigl(1-
\bigl(\lambda^{(\varrho)}1_{\auf 0,\sigma\zu}
+\bigl(1-\lambda^{(\varrho)}\mal S_\sigma\bigr)\vartheta \bigr) \mal
S_T \bigr)^2 |\FFF_\sigma \bigr)
\end{eqnarray}
holds almost surely for any $\vartheta\in\overline\Theta$ with
$\vartheta 1_{\auf 0,\sigma\zu}=0$. Otherwise, there exists some
$\vartheta\in\overline\Theta$ with $\vartheta 1_{\auf 0,\sigma\zu}=0$
such that the reverse inequality holds on some set $F\in\FFF_\sigma$
with $P(F)>0$. Define the strategy
\[
\psi:= \cases{ \lambda^{(\varrho)}1_{\auf0,\sigma\zu}
+\bigl(1-\lambda^{(\varrho)}\mal S_\sigma\bigr)\vartheta, &\quad
$\mbox{on }F$, \cr\noalign{} \lambda^{(\varrho)},&\quad
$\mbox{on }F^C$.}
\]
We have
\begin{eqnarray}\label{e:wider}
\nonumber && E\bigl((1-\psi\mal S_T)^2\bigr)
\\
&&\qquad = E \bigl(E \bigl( \bigl(1-\lambda^{(\varrho)}\mal S_T\bigr)^2
|\FFF _\sigma \bigr)1_{F^C} \bigr) \nonumber
\\[-8pt]
\\[-8pt]
\nonumber &&\qquad\quad{}+E \bigl(E \bigl(  \bigl(1-
\bigl(\lambda^{(\varrho )}1_{\auf 0,\sigma\zu}
+\bigl(1-\lambda^{(\varrho)}\mal S_\sigma\bigr)\vartheta \bigr)\mal S_T
\bigr)^2 |\FFF_\sigma \bigr)1_F \bigr)
\\
\nonumber &&\qquad < E \bigl(\bigl(1-\lambda^{(\varrho)}\mal
S_T\bigr)^2 \bigr).
\end{eqnarray}
This contradicts the optimality of $\lambda^{(\varrho)}$ if $\psi\in
\overline \Theta$.

In order to show $\psi\in\overline\Theta$, let $Z$ be the density
process of some S$\sigma$MM with square-integrable density. Integration
by parts yields that $(\psi\mal S)Z$ is a \mbox{$\sigma$-}martingale
[cf. (\ref{e:partial})]. Since $P(|\lambda^{(\varrho)}\mal S_\sigma|\le
n)\uparrow1$ and $P(|\vartheta\mal S_\sigma|\le n)\uparrow1$ for
$n\uparrow\infty$, we may assume w.l.o.g. that
$|\lambda^{(\varrho)}\mal S_\sigma|$ and $|\vartheta\mal S_\sigma|$ are
bounded on $F$, say by $n\in\nn$. On $\zu\sigma,T\zu$ we have
\begin{eqnarray*}
&& \big|\bigl(\psi\mal S-\lambda^{(\varrho)}\mal S\bigr)Z \big|
\\
&&\qquad \leq \bigl(\big|\lambda^{(\varrho)}\mal
S-\lambda^{(\varrho)}\mal S_\sigma\big| +\big|1-\lambda^{(\varrho)}\mal
S_\sigma\big||\vartheta\mal S-\vartheta\mal S_\sigma | \bigr)|Z| 1_F
\\
&&\qquad \leq \bigl(\big|\bigl(\lambda^{(\varrho)}\mal
S\bigr)Z\big|+n|Z| +(n+1) \bigl(|(\vartheta\mal S) Z|+n \bigr) \bigr)
1_F.
\end{eqnarray*}
The processes in the last line are of class (D) by Corollary
\ref{co:theta}. This in turn implies that $(\psi\mal S)Z$ is of class
(D) as well and hence a martingale. Another application of Corollary
\ref{co:theta} yields $\psi\in \overline \Theta$. Thus (\ref{e:wider})
yields a true contradiction, which means that (\ref{e:kleiner}) holds.

Note that (\ref{e:kleiner}) implies
\begin{equation}\label{e:kleiner2}
E \biggl(  \biggl(1-{\lambda^{(\varrho)}1_{\zu\sigma,T\zu}\over
1-\lambda^{(\varrho)}\mal S_\sigma}\mal S_T \biggr)^2 \Big|\FFF_\sigma
 \biggr)\leq E \bigl(  (1-\vartheta\mal S_T )^2 |\FFF _\sigma
 \bigr)
\end{equation}
almost surely on $\{1-\lambda^{(\varrho)}\mal S_\sigma\not=0\}$ for any
$\vartheta\in\overline\Theta$ with $\vartheta1_{\auf0,\sigma \zu}=0$.
Moreover, we have on the set $\{1-\lambda^{(\varrho)}\mal S_\sigma=0\}$
that
\[
E \bigl( \bigl(1-\lambda^{(\varrho)}\mal S_T\bigr)^2 |\FFF_\sigma
 \bigr)
\leq E \bigl(
\bigl(1-\bigl(\lambda^{(\varrho)}1_{\auf0,\sigma\zu}\bigr)\mal
S_T\bigr)^2 |\FFF_\sigma \bigr)=0
\]
and hence $1-\lambda^{(\varrho)}\mal S_T=0$.

Similarly as (\ref{e:kleiner}), one shows that
\begin{equation}\label{e:kleiner3}
\hspace*{6mm} E \bigl( \bigl(1-\lambda^{(\sigma)}\mal S_T\bigr)^2
\big|\FFF_\sigma
 \bigr)
\leq E \biggl(  \biggl(1- \biggl(\alpha{\lambda^{(\varrho)}1_{\zu
\sigma ,T\zu} \over1-\lambda^{(\varrho)}\mal S_\sigma}+\vartheta
\biggr)\mal S_T \biggr)^2 \Big|\FFF_\sigma \biggr)
\end{equation}
holds almost surely on $\{1-\lambda^{(\varrho)}\mal S_\sigma\not=0\}$
for any $\alpha\in\rp$ and any $\vartheta\in\overline\Theta$ with
\mbox{$\vartheta1_{\auf0,\sigma\zu}=0$}. Using a convexity argument,
(\ref{e:kleiner2}) and (\ref{e:kleiner3}) yield that
\[
1-\lambda^{(\sigma)}\mal S_T=1-{\lambda^{(\varrho)}1_{\zu\sigma,T\zu}
\over1-\lambda^{(\varrho)}\mal S_\sigma}\mal S_T
\]
on $\{1-\lambda^{(\varrho)}\mal S_\sigma\not=0\}$ and hence
\[
\lambda^{(\sigma)}\mal S_T\bigl(1-\lambda^{(\varrho)}\mal
S_\sigma\bigr) =\bigl(\lambda^{(\varrho)}1_{\zu\sigma,T\zu}\bigr)\mal
S_T.
\]
By taking conditional expectation relative to the $\sigma$-martingale
measure $Q$ from Assumption \ref{a:eins}, it follows that
\[
\lambda^{(\sigma)}\mal S_\tau\bigl(1-\lambda^{(\varrho)}\mal
S_\sigma\bigr) =\bigl(\lambda^{(\varrho)}1_{\zu\sigma,T\zu}\bigr)\mal
S_\tau
\]
for any $\tau\geq\sigma$ (cf. Corollary \ref{co:theta}), which yields
the claim.

\item[3.] This is shown in the proof of statement 2.

\item[4.] This follows from (\ref{e:kleiner}) for $\varrho=\sigma$.

\item[5.] If $E((1-\lambda^{(\tau)}\mal S_T)^2|\FFF_\sigma)=0$ on some
set $F\in\FFF _\sigma$ with $P(F)>0$, then
\[
\lambda^{(\tau)}\mal S_T-\lambda^{(\tau)}\mal S_\sigma=1,
\]
which contradicts the fact that $\lambda^{(\tau)}\mal S$ is a
$Q$-martingale for the $\sigma $-martingale measure $Q$ from Assumption
\ref{a:eins} (cf. Corollary \ref{co:theta}). Hence,
$E((1-\lambda^{(\tau)}\mal S_T)^2| \FFF_\sigma)>0$ almost surely.
Moreover,
\begin{eqnarray*}
E \bigl( \bigl(1-(1+\varepsilon)\lambda^{(\tau)}\mal S_T\bigr)^2
|\FFF_{\sigma}
\bigr)&=&  E \bigl( \bigl(1-\lambda^{(\tau)}\mal S_T\bigr)^2 |\FFF_{\sigma } \bigr)
\\
&&{}-2\varepsilon E \bigl( \lambda^{(\tau)}\mal S_T
\bigl(1-\lambda^{(\tau)}\mal S_T\bigr) |\FFF_{\sigma} \bigr)
\\
&&{}  +\varepsilon^2E \bigl( \bigl(\lambda^{(\tau)}\mal S_T\bigr)^2
|\FFF _{\sigma } \bigr)
\end{eqnarray*}
for any $\varepsilon\in\rr$. By statement 4 this implies
$E(\lambda^{(\tau)}\mal S_T(1-\lambda^{(\tau)}\mal
S_T)|\FFF_{\sigma})=0$. Together, the assertion follows.\quad\qed
\end{longlist}}\noqed
\end{pf}

\begin{lemma}\label{l:L}
\textup{1.} There exists a unique semimartingale $L$ with $L_T=1$ such
that the process $M^{(\tau)}-(M^{(\tau)})^\tau$ is a martingale for any
stopping time $\tau$, where
\begin{equation}\label{e:mtau}
M^{(\tau)}:=\bigl(1-\lambda^{(\tau)}\mal S\bigr)L.
\end{equation}

{\smallskipamount=0pt
\begin{longlist}[2.]
\item[2.] The process $1_{\zu\tau,T\zu}\mal(SM^{(\tau)})$ is a
martingale for any stopping time $\tau$. (In the slightly more general
setup of Remark~\textup{\ref{r:general}}, the upper bound $T$ is to be
replaced by $U_n$ for arbitrary $n$.)

\item[3.] The process $((v+\vartheta\mal S_s)M^{(t)}_s)_{s\in[t,T]}$ is
a martingale for any
$(v,\vartheta)\in\overline{L^2(\FFF_0)\times\Theta}$ and any
$t\in[0,T]$.
\end{longlist}}
\end{lemma}

\begin{pf}
1. Our reasoning relies heavily on the proofs of Lemma~3.4 and
Theorem~1.3 in \cite{delbaenschachermayer96b}. For any stopping time
$\sigma$ we introduce the process
\[
{}^\sigma M_t:= {E(1-\lambda^{(\sigma)}\mal S_T|\FFF_t)\over
E(1-\lambda^{(\sigma)}\mal S_T|\FFF_{\sigma\wedge t})}.
\]
Define stopping times $(\tau_n)_{n\in\nn}$ recursively by $\tau
_0:=0$ and
\[
\tau_{n+1}:=\inf \biggl\{t\geq\tau_n\dvtx
 \bigg|{1-\lambda^{(\tau_n)}\mal S_t\over E(1-\lambda^{(\tau
_n)}\mal S_T|\FFF_{\tau_n})} \bigg| \leq{1\over2} \biggr\}\wedge T.
\]
Then
\[
|{}^{\tau_n}M_{\tau_{n+1}}|= {|1-\lambda^{(\tau_n)}\mal S_{\tau_{n+1}}|
\over E(1-\lambda^{(\tau_n)}\mal S_T|\FFF_{\tau_n})}
\big|E\bigl(1-\lambda^{(\tau_{n+1})}\mal S_T|\FFF_{\tau_{n+1}}\bigr)
\big|\leq {1\over2}
\]
on $\{\tau_{n+1}<T\}$ by Lemma~\ref{l:existenz}. Using
Lemma~\ref{l:existenz}(2) one easily verifies that
\[
{}^{\tau_n}M_t={}^{\tau_n}M_{\tau_{n+1}}{}^{\tau_{n+1}}M_t
\]
for $t\geq\tau_{n+1}$.
Consequently, $\lim_{m\to\infty}{}^{\tau_n}M_{\tau_m}=0$
on $D:=\{\tau_n<T$ for all $n\in\nn\}$.
Letting
\[
\widetilde M^{(\tau_n)}_t:=E\bigl(1-\lambda^{(\tau_n)}\mal
S_T|\FFF_t\bigr)
\]
we have
\begin{eqnarray*}
1&=& \lim_{m\to\infty}{E(\widetilde M^{(\tau_n)}_{\tau_m}|\FFF_{\tau
_n})\over \widetilde M^{(\tau_n)}_{\tau_n}}
\\
&=&E \biggl( {\lim_{m\to\infty}\widetilde M^{(\tau_n)}_{\tau _m}\over
\widetilde M^{(\tau_n)}_{\tau_n}} \Big|\FFF_{\tau_n} \biggr)
\\
&=&E \biggl(
\lim_{m\to\infty}{}^{\tau_n}M_{\tau_m} \Big|\FFF_{\tau _n} \biggr)
\\
&=&E({}^{\tau_n}M_T1_{D^C}|\FFF_{\tau_n})
\\
&\leq&\sqrt{E(({}^{\tau_n}M_T)^2|\FFF_{\tau_n})}\sqrt{E(1_{D^C}|\FFF
_{\tau_n})}.
\end{eqnarray*}
Since the last term converges to 0 on $D$, it follows that
\begin{equation}\label{e:onD}
\lim_{n\to\infty}{E(({}^{\tau_n}M_T)^2|\FFF_{\tau_n})}=\infty\qquad\mbox{on
}D.
\end{equation}

Denote by $Z$ the density process of the measure $Q$ from Assumption
\ref{a:eins}.
By Corollary \ref{co:theta} we have
\begin{equation}\label{e:zdurchz}
E \biggl( \bigl(1-\lambda^{(\tau_n)}\mal S_T\bigr) {Z_T\over
Z_{\tau_n}} \Big|\FFF_{\tau_n} \biggr)=1.
\end{equation}
Observe that
\begin{eqnarray*}
E \biggl(  \biggl({Z_T\over Z_{\tau_n}} \biggr)^2 \Big|\FFF _{\tau _n}
\biggr) &=&E ( ({}^{\tau_n}M_T)^2 |\FFF_{\tau_n} ) +2E \biggl(
{}^{\tau_n}M_T
 \biggl({Z_T\over Z_{\tau_n}}-{}^{\tau_n}M_T \biggr) \bigg|\FFF_{\tau
_n} \biggr)
\\
&&{}+E \biggl(  \biggl({Z_T\over Z_{\tau_n}}-{}^{\tau _n}M_T \biggr)^2
\Big|\FFF_{\tau_n} \biggr)
\end{eqnarray*}
for any $n\geq1$. Due to (\ref{e:zdurchz}) and
Lemma~\ref{l:existenz}(5) the second term on the right-hand side
vanishes. It follows that
\begin{equation}\label{e:step2}
E ( ({}^{\tau_n}M_T)^2 |\FFF_{\tau_n} ) \leq E \biggl( \biggl({Z_T\over
Z_{\tau_n}} \biggr)^2 \Big|\FFF _{\tau _n} \biggr).
\end{equation}

Together we have $P(D)=0$: Indeed, otherwise (\ref{e:onD}) yields
\[
P \biggl(\,\sup_{t\in[0,T]}{E(Z_T^2|\FFF_t)\over Z_t^2}
<E(({}^{\tau_n}M_T)^2|\FFF_{\tau_n}) \biggr)>0
\]
for large $n$. Consequently,
\[
 \biggl\{{E(Z_T^2|\FFF_{\tau_n})\over Z_{\tau_n}^2}<E(({}^{\tau
_n}M_T)^2|\FFF _{\tau_n}) \biggr\} \in\FFF_{\tau_n}
\]
has positive probability as well in contradiction to (\ref{e:step2}).

Now define the semimartingale $L$ by
\[
L_t:={E(1-\lambda^{(\tau_n)}\mal S_T|\FFF_t)\over1-\lambda^{(\tau
_n)}\mal S_t}\qquad \mbox{for }\tau_{n}\leq t<\tau_{n+1}.
\]
The claimed martingale property follows from Lemma~\ref{l:existenz}(2).

Uniqueness of $L$ follows from
\[
E\bigl(1-\lambda^{(t)}\mal
S_T|\FFF_t\bigr)-L_t=E\bigl(M^{(t)}_T|\FFF_t\bigr)-M^{(t)}_t=0.
\]
{\smallskipamount=0pt
\begin{longlist}[2.]
\item[2.] It suffices to verify that $E(1_{\zu\tau,T\zu}\mal(SM^{(\tau
)})_\sigma)=0$ for any stopping time $\sigma$. By substituting
$\sigma\vee\tau$ for $\sigma$, we may assume $\sigma\geq\tau$ w.l.o.g.
Since $1_{\zu\tau,T\zu}\mal M^{(\tau)}$ is a square-integrable
martingale, we have $E(S_\sigma(M^{(\tau)}_T-M^{(\tau)}_\sigma))=0$ and
similarly $E(S_\tau(M^{(\tau)}_T-M^{(\tau)}_\tau))=0$. Consequently,
\begin{eqnarray*}
E \bigl(S_\sigma M^{(\tau)}_\sigma-S_{\tau} M^{(\tau)}_{\tau}
 \bigr)
=E \bigl((S_\sigma-S_{\tau})M^{(\tau)}_T \bigr) =E \bigl((\psi\mal
S_T)M^{(\tau)}_T \bigr)
\end{eqnarray*}
for $\psi:=1_{\zu\tau,\sigma\zu}$.
The optimality of $\lambda^{(\tau)}$ implies that
\begin{eqnarray*}
0&\leq&E \bigl(\bigl(1-\bigl(\lambda^{(\tau)}+\varepsilon\psi\bigr)\mal
S_T\bigr)^2 \bigr)- E \bigl(\bigl(1-\lambda^{(\tau)}\mal S_T\bigr)^2 \bigr)
\\
&=& 2\varepsilon E \bigl((\psi\mal S_T)M^{(\tau)}_T \bigr)
+\varepsilon^2 E \bigl((\psi\mal S_T)^2 \bigr)
\end{eqnarray*}
for any $\varepsilon\in\rr$ and hence
$E((\psi\mal S_T)M^{(\tau)}_T)=0$.

\item[3.] By statement 2 we have that $1_{\zu t,T\zu}\mal(SM^{(t)})$
and hence $(S_sM^{(t)}_s)_{s\in[t,T]}$ is a martingale. Consequently,
the signed measure with density process $(M^{(t)}/E(L_t))_{s\in[t,T]}$ is a S$\sigma$MM in the sense of
Definition~\ref{d:SSMM} if the time set $[0,T]$ is replaced with
$[t,T]$. By Lemma~\ref{l:k2f0} (also adapted to $[t,T]$ instead of
$[0,T]$ as time set), the assertion follows.\quad\qed
\end{longlist}}\noqed
\end{pf}

\begin{defi}\label{d:L}
We call the process $L$ from Lemma~\ref{l:L} \textit{opportunity
process}.
\end{defi}

The terminology is inspired by the fact that $L$ is linked to optimal
investment opportunities.
Indeed, the following corollary states that $L$ represents both first
and second moments of
efficient strategies in the sense of (\ref{e:efficient}).

\begin{cor}\label{co:L}
For any $t\in[0,T]$ we have
\begin{eqnarray}\label{e:infimum}
\nonumber L_t&=&E\bigl(1-\lambda^{(t)}\mal S_T|\FFF_t\bigr)
\\
&=&E \bigl( \bigl(1-\lambda^{(t)}\mal S_T\bigr)^2 |\FFF_t \bigr)
\\
\nonumber &=&\inf \bigl\{E \bigl( (1-\vartheta\mal S_T)^2 |\FFF _t
\bigr)\dvtx \vartheta\in\overline\Theta\mbox{ with }
\vartheta1_{\auf0,t\zu}=0 \bigr\}.
\end{eqnarray}
In particular, $L$ is a submartingale.
\end{cor}

\begin{pf}This follows from Lemmas \ref{l:L} and
\ref{l:existenz}.
\end{pf}

These equations can be interpreted in terms of dynamic Sharpe ratios
(cf. also \cite{leitner01}, (5.16)):

\begin{defi}
For $t\in[0,T]$ we call
\begin{equation}\label{e:sharpe}
\varrho_t:=\sup \biggl\{{E(\vartheta\mal S_T|\FFF_t)\over\sqrt{\Var
(\vartheta\mal S_T|\FFF_t)}}\dvtx  \vartheta\in\overline\Theta\mbox{
with }\vartheta1_{\auf0,t\zu }=0 \biggr\}
\end{equation}
\textit{maximal Sharpe ratio on $(t,T]$}, where we set
$\Var(X|\FFF_t):=E(X^2|\FFF_t)-(E(X|\FFF_t))^2$ and ${0\over0}:=0$.
\end{defi}

\begin{prop}\label{p:sharpe}
The relation between opportunity process $L$ and maximal Sharpe ratio
$\varrho$ is given by
\[
\varrho=\sqrt{{1\over L}-1}
\]
and
\[
L={1\over1+\varrho^2},
\]
respectively.
\end{prop}

\begin{pf}On the set
\[
D:= \bigl\{\omega\in\Omega\dvtx E(\vartheta\mal
S_T|\FFF_t)(\omega)=0\mbox { for all }
\vartheta\in\overline\Theta\mbox{ with }\vartheta1_{\auf0,t\zu }=0
\bigr\}
\]
we have $\varrho_t=0$. Moreover, the infimum in (\ref{e:infimum}) is
attained in $\lambda^{(t)}=0$, which implies that $L_t=1$ on $D$.

For $\omega\in D^C$ there exists some $\vartheta\in\overline\Theta$
with $\vartheta 1_{\auf0,t\zu=0}$ and $E(\vartheta\mal
S_T|\FFF_t)(\omega)>0$. For sufficiently small $\varepsilon>0$ we have
that $E((1-\varepsilon\vartheta\mal S_T)^2|\FFF_t)(\omega)<1$, which
implies that $L_t<1$ on $D^C$ (cf.  Corollary \ref{co:L}). By scaling
invariance it suffices to consider $\vartheta$ with $E(\vartheta\mal
S_T|\FFF_t)=1-L_t$ in the supremum of (\ref{e:sharpe}). For these
$\vartheta$ we have
\[
{E(\vartheta\mal S_T|\FFF_t)\over\sqrt{\Var(\vartheta\mal S_T|\FFF_t)}}
={1-L_t\over\sqrt{E((1-\vartheta\mal S_T)^2|\FFF_t)-L_t^2}},
\]
which implies that the supremum is attained in
$\vartheta=\lambda^{(t)}$. The assertion follows now from Corollary
\ref{co:L}.
\end{pf}

\subsection{\texorpdfstring{Adjustment process.}{Adjustment process}}
The optimal number of shares $\lambda^{(\tau)}$ in (\ref{e:efficient})
depends on $\tau$. However, the optimal number of shares \textit{per
unit of wealth} does not. It is denoted by $\tilde a$ in the following
lemma.

\begin{lemma}\label{l:exa}
We use the notation from Lemma~\textup{\ref{l:existenz}}. There exists
some $\tilde a\in L(S)$ such that
\[
1-\lambda^{(\tau)}\mal S =\EEE \bigl(\bigl(-\tilde
a1_{\zu\tau,T\zu}\bigr)\mal S \bigr) =1- \bigl(\tilde
a1_{\zu\tau,T\zu}\EEE\bigl(\bigl(-\tilde a1_{\zu\tau,T\zu }\bigr)\mal
S\bigr)_- \bigr)\mal S
\]
for any stopping time $\tau$. Consequently, we may assume
\begin{equation}\label{e:mayassume}
\lambda^{(\tau)}=\tilde a1_{\zu\tau,T\zu}\EEE\bigl(\bigl(-\tilde
a1_{\zu\tau ,T\zu }\bigr)\mal S\bigr)_-.
\end{equation}
\end{lemma}

\begin{pf}
Let
\[
\tilde a:=\sum_{n=0}^\infty
{\lambda^{(\tau_n)}\over1-\lambda^{(\tau_n)}\mal S_-} 1_{\zu\tau
_{n},\tau_{n+1}\zu},
\]
where $(\tau_n)_{n\in\nn}$ denotes the sequence of stopping times from
the proof of Lemma~\ref{l:L}. On $\auf0,\tau_{n+1}\zu$ we have
\[
1-\lambda^{(\tau_n)}\mal S=1- \bigl(\bigl(1-\lambda^{(\tau_n)}\mal
S_-\bigr) \tilde a1_{\zu\tau_{n},T\zu} \bigr)\mal S
\]
and hence
\begin{equation}\label{e:gleich}
1-\lambda^{(\tau_n)}\mal S=\EEE\bigl(\bigl(-\tilde
a1_{\zu\tau_{n},T\zu}\bigr)\mal S\bigr).
\end{equation}
From
\[
1-\lambda^{(\tau_n)}\mal S_t =\bigl(1-\lambda^{(\tau_n)}\mal
S_{\tau_{n+1}}\bigr)\bigl(1-\lambda^{(\tau _{n+1})}\mal S_t\bigr)
\]
and
\[
\EEE \bigl(\bigl(-\tilde a1_{\zu\tau_{n},T\zu}\bigr)\mal S \bigr)_t
=\EEE \bigl(\bigl(-\tilde a1_{\zu\tau_{n},T\zu}\bigr)\mal S
\bigr)_{\tau_{n+1}} \EEE \bigl(\bigl(-\tilde
a1_{\zu\tau_{n+1},T\zu}\bigr)\mal S \bigr)_t
\]
for $t\in\zu\tau_{n+1},\tau_{n+2}\zu$
it follows recursively that (\ref{e:gleich}) holds on $[0,T]$.
Now let $\tau$ be arbitrary. On $\{\tau_n\leq\tau<\tau_{n+1}\}$ we have
\[
1-\lambda^{(\tau)}\mal S ={1-\lambda^{(\tau_n)}\mal
S\over1-\lambda^{(\tau_n)}\mal S_\tau} ={\EEE ((-\tilde
a1_{\zu\tau_{n},T\zu})\mal S )\over \EEE ((-\tilde
a1_{\zu\tau_{n},T\zu})\mal S )_\tau} =\EEE \bigl(\bigl(-\tilde
a1_{\zu\tau,T\zu}\bigr)\mal S \bigr)
\]
as claimed.
\end{pf}

\begin{defi}\label{d:moda}
The (not necessarily unique) process $\tilde a$ from Lemma~\ref{l:exa}
is called \textit{adjustment process}. Moreover, we call
\[
\hat a:=(1+\Delta A^K)\tilde a
\]
\textit{extended adjustment process}.
\end{defi}

The name \textit{adjustment process} is taken from \cite{schweizer96}:

\begin{cor}\label{co:adjustment}
$E(\vartheta\mal S_T\EEE(-\tilde a \mal S)_T)=0$ for any $\vartheta\in
\overline \Theta $, i.e., $\tilde a$ is an adjustment process in the
sense of \textup{\cite{schweizer96}}, Section~\textup{3} with
$\overline\Theta$ substituted for $\Theta$.
\end{cor}

\begin{pf}
This follows from Lemma~\ref{l:L}(3).
\end{pf}

\begin{lemma}\label{l:values}
$L,L_-$ are $(0,1]$-valued.
\end{lemma}
\begin{pf}Lemma~\ref{l:existenz}(5) implies that
$L_t=E(1-\lambda^{(t)}\mal S_T|\FFF_t)\in(0,1]$ almost surely for fixed
$t$, which yields by right-continuity that $L$ is $[0,1]$-valued
outside some evanescent set.

Let $\tau:=\inf\{t\in[0,T]:L_t=0\}\wedge T$. Again by
Lemma~\ref{l:existenz}(5), we have
\[
0=L_{\tau\wedge T}= E \bigl( \bigl(1-\lambda^{(\tau\wedge T)}\mal
S_T\bigr)^2 |\FFF _{\tau \wedge T} \bigr)\in(0,1]
\]
on $\{L_t=0$ for some $t\in[0,T]\}$, which implies that
\begin{equation}\label{e:wkeit0}
P(L_t=0\mbox{ for some }t\in[0,T])=0.
\end{equation}

Finally let $\tau:=\inf\{t\in[0,T]\dvtx  L_{t-}=0\}\wedge T$. Define an
increasing sequence of stopping times $(\tau_n)_{n\in\nn}$ via
$\tau_n:=\inf\{t\in[0,T]\dvtx L_t\leq{1\over n}\}\wedge T$. By
(\ref{e:wkeit0}) we have $\tau_n\uparrow\uparrow\tau$ on $\{ L_{\tau
-}=0\}$. Lemma~\ref{l:existenz}(5) implies
\[
E \bigl(\bigl(1-\lambda^{(\tau_n)}\mal S_T\bigr)^21_{\{\tau_n<T\}}
\bigr) =E \bigl(L_{\tau_n}1_{\{\tau_n<T\}} \bigr).
\]
By \cite{protter04}, Theorem~V.13 we have that
\[
1-\lambda^{(\tau_n)}\mal S_T=\EEE \bigl(\bigl(-\tilde a1_{\zu\tau
_n,T\zu }\bigr)\mal S \bigr)_T \to\EEE \bigl(\bigl(-\tilde
a1_{\auf\tau,T\zu\cap\{L_{\tau-}=0\} }\bigr)\mal S \bigr)_T
\]
in probability for $n\to\infty$. In view of Fatou's lemma and dominated
convergence, we obtain
\[
0\leq E \bigl( \bigl(\EEE\bigl(\bigl(-\tilde
a1_{\auf\tau,T\zu}\bigr)\mal S\bigr)_T \bigr)^21_{\{L_{\tau-}=0\}}
\bigr) \leq E \bigl(L_{\tau-}1_{\{L_{\tau-}=0\}} \bigr)=0.
\]
Suppose that $\{L_-=0\}$ is not evanescent. Then there is some $n$ such that
$P(D)>0$ for
\begin{eqnarray*}
D &:=& \{L_{\tau-}=0\mbox{ and }1-\lambda^{(\tau_n)}\mal S_{\tau -}>0
\}
\\
&\hspace*{3pt} \supset& \{L_{\tau-}=0\mbox{ and }\Delta(-\tilde a\mal
S)>-1\mbox { on }\zu\tau_n,\tau\auf \}.
\end{eqnarray*}
On $D\in\FFF_{\tau-}$ we have
\[
{1-\lambda^{(\tau_n)}\mal S_T\over1-\lambda^{(\tau_n)}\mal S_{\tau-}}=
{\EEE ((-\tilde a1_{\zu\tau_n,T\zu})\mal S )_T\over \EEE ((-\tilde
a1_{\zu\tau_n,T\zu})\mal S )_{\tau-}}= \EEE \bigl(\bigl(-\tilde
a1_{\auf\tau,T\zu}\bigr)\mal S \bigr)_T=0.
\]
Consequently, the process $\lambda^{(\tau_n)}\mal S$ cannot be a
martingale under the $\sigma$-martingale measure from Assumption
\ref{a:eins}, which yields a contradiction to Corollary~\ref{co:theta}.
\end{pf}

Since $L_-$ does not vanish, the stochastic logarithm of $L$ is well defined:
\begin{defi}\label{d:K}
We call
\[
K:=\LLL(L):={1\over{L_-}}\mal L
\]
\textit{modified mean-variance tradeoff \textup{(}MMVT\textup{)}
process}.
\end{defi}

The modified mean-variance tradeoff process is related to the
\textit{mean-variance tradeoff \textup{(}MVT)\textup{} process} of
\cite{schweizer94} (cf. Section \ref{su:ppstar}).

\subsection{\texorpdfstring{Variance-optimal signed martingale
measure.}{Variance-optimal signed martingale measure}} With the help of
the modified mean-variance tradeoff process $K$ and the adjustment
process $\tilde a$ we can define a signed measure $Q^\star$ which plays
an important role in the context of quadratic hedging. This
\textit{variance-optimal signed martingale measure} appears more or
less explicitly in many papers on the subject.

\begin{defi}\label{d:N}
We call
\begin{equation}\label{e:N}
N:=K-\tilde a\mal S-[\tilde a\mal S,K]
\end{equation}
\textit{variance-optimal logarithm process} and the signed measure
$Q^\star$ defined via
\begin{equation}\label{e:VOMM}
{dQ^\star\over dP}:={L_0\over E(L_0)}\EEE(N)_T={\EEE(-\tilde a\mal
S)_T\over E(L_0)} ={1-\lambda^{(0)}\mal S_T\over E(1-\lambda^{(0)}\mal
S_T)}
\end{equation}
\textit{variance-optimal signed martingale measure
\textup{(}variance-optimal S$\sigma$MM\textup{)}}.
\end{defi}

The following result explains the terminology.

\begin{prop}\label{p:VOMM}
\textup{1.} $Q^\star$ is a S$\sigma$MM (cf. Definition~\textup{\ref{d:SSMM}})
with density process
\[
Z^{Q^\star}:={L_0\over E(L_0)}\EEE(N)={L\EEE(-\tilde a\mal S)\over
E(L_0)}.
\]

\textup{2.} $Q^\star$ minimizes $Q\mapsto E(({dQ\over dP})^2)$ over all
S$\sigma$MM\textup{'}s $Q$. Hence it is the \textit{variance-optimal
signed $\overline\Theta$-martingale measure} in the sense of
\textup{\cite{schweizer96}}, Section~\textup{1}, with $\Theta$ replaced
by $\overline \Theta$ in the definition.
\end{prop}

\begin{pf}
\textup{1.} Note that $L_0\EEE(N)=M^{(0)}$ is a martingale by
Lemma~\ref{l:L}. Lemma~\ref{l:L}(3) implies that $Q^\star$ is a
S$\sigma$MM.

2. For any other S$\sigma$MM $Q$ with ${dQ\over dP}\in L^2(P)$ we have
\begin{eqnarray*}
&& E \biggl( \biggl({dQ\over dP} \biggr)^2 \biggr) -E \biggl(
\biggl({dQ^\star\over dP} \biggr)^2 \biggr)
\\
&&\qquad  \geq 2E \biggl( \biggl({dQ\over dP}-{dQ^\star\over dP}
\biggr){dQ^\star \over dP} \biggr)
\\
&&\qquad = 2E \biggl({dQ\over dP}{1-\lambda^{(0)}\mal S_T\over E(L_0)}
\biggr)- 2E \biggl({dQ^\star\over dP}{1-\lambda^{(0)}\mal S_T\over
E(L_0)} \biggr)=0
\end{eqnarray*}
by Corollary \ref{co:theta}.
\end{pf}

If $Q\sim P$ is a probability measure with density process $Z=\EEE(M)$,
then the density ${dQ\over dP}$, the density process $Z$, and its
stochastic logarithm $M$ uniquely determine one another.
This is not true for the variance-optimal S$\sigma$MM 
$Q^\star$ because $\EEE(N)$ may vanish and hence $N$ cannot be fully
recovered from $\EEE(N)$ or ${dQ^\star\over dP}$. Therefore the
following result does not follow immediately from the fact that
$Q^\star$ is a S$\sigma$MM whose density process is a multiple of
$\EEE(N)$.

\begin{lemma}\label{l:N}
The variance-optimal logarithm process $N$ and also $S+[S,N]$ are
$\sigma$-mar\-tingales. Consequently, $S\EEE(N)$ is a
$\sigma$-martingale as well.
\end{lemma}

\begin{pf}Denote by $(\tau_n)_{n\in\nn}$ the sequence of
stopping times from the proof of Lemma~\ref{l:L}. Since
$\bigcup_{n\in\nn}\zu\tau_n,\tau_{n+1}\zu=\Omega\times(0,T]$, it
suffices to show that \mbox{$1_{\zu\tau_n,\tau_{n+1}\zu}\mal N$} and
$1_{\zu\tau_n,\tau_{n+1}\zu}\mal(S+[N,S])$ are $\sigma$-martingales for
any $n\in\nn$. Since
\begin{eqnarray*}
\EEE(N-N^{\tau_n})&=& \EEE \bigl(1_{\zu\tau_n,T\zu}\mal(K-\tilde a\mal
S-[\tilde a\mal S,K]) \bigr)
\\
&=&\EEE \bigl(1_{\zu\tau_n,T\zu}\mal K \bigr) \EEE
\bigl(\bigl(-\tilde a1_{\zu\tau_n,T\zu}\bigr)\mal S \bigr)
\\
&=&{L(1-\lambda^{(\tau_n)}\mal S)\over L^{\tau_n}}
\\
&=&1+{1_{\zu\tau_n,T\zu}\over L_{\tau_n}}\mal M^{(\tau_n)}
\end{eqnarray*}
is a $\sigma$-martingale, we have that
\[
1_{\zu\tau_n,\tau_{n+1}\zu}\mal N
={1_{\zu\tau_n,\tau_{n+1}\zu}\over\EEE(N-N^{\tau_n})_-}\mal\EEE
(N-N^{\tau_n})
\]
is a $\sigma$-martingale as well.
Similarly,
\begin{eqnarray*}
1_{\zu\tau_n,T\zu}\mal\bigl(\EEE(N-N^{\tau_n})S\bigr)
&=&1_{\zu\tau_n,T\zu}\mal{L(1-\lambda^{(\tau_n)}\mal S)S\over L^{\tau
_n}}
\\
&=&{1_{\zu\tau_n,T\zu}\over L_{\tau_n}}\mal \bigl(1_{\zu\tau _n,T\zu}
\mal\bigl(M^{(\tau_n)}S\bigr) \bigr)
\end{eqnarray*}
is a $\sigma$-martingale by Lemma~\ref{l:L}(2). Integration by parts
yields
\begin{eqnarray*}
&& 1_{\zu\tau_n,T\zu}\mal \bigl(\EEE(N-N^{\tau_n})S \bigr)
-S_-\mal\EEE(N-N^{\tau_n})
\\
&&\qquad =
\bigl(\EEE(N-N^{\tau_n})_-1_{\zu\tau_n,T\zu} \bigr)\mal(S+[N,S]),
\end{eqnarray*}
which implies that
\begin{eqnarray*}
&& 1_{\zu\tau_n,\tau_{n+1}\zu}\mal(S+[N,S])
\\
&&\qquad
={1_{\zu\tau_n,\tau_{n+1}\zu}\over\EEE(N-N^{\tau_n})_-}\mal
 \bigl(\bigl(\EEE(N-N^{\tau_n})_-1_{\zu\tau_n,T\zu}\bigr)\mal(S+[N,S]) \bigr)
\end{eqnarray*}
is a $\sigma$-martingale as well.
Finally,
\[
S\EEE(N)=S_0+\EEE(N)_-\mal(S_-\mal N+S+[S,N])
\]
yields the last assertion.
\end{pf}

\subsection{\texorpdfstring{Opportunity-neutral measure.}{Opportunity-neutral
measure}} In this section we define a measure $P^\star$ in terms of its
density process
\[
Z^{P^\star}:={L\over E(L_0)\EEE(A^K)}.
\]
For $Z^{P^\star}$ to be truly a density process, we need the
following
\begin{lemma}\label{l:K}
The process
$Z^{P^\star}$ is a bounded positive martingale and satisfies
\[
Z^{P^\star}={L_0\over E(L_0)}\EEE \biggl({1\over1+\Delta A^K}\mal M^K
\biggr).
\]
\end{lemma}

\begin{pf}
Since $L$ is a submartingale by Corollary \ref{co:L}, we have $b^L\geq
0$ and hence $b^K={1\over L_-}b^L\geq0$ outside some
$P\otimes A$-null set. This implies that \mbox{$A^K=b^K\mal A$} and
hence also $\EEE(A^K)$ are increasing processes. Thus we have
$0<Z^{P^\star}\leq{1\over E(L_0)}$. The equality of the two expressions
for $Z^{P^\star}$ follows from Yor's formula. From the second
representation we conclude that $Z^{P^\star}$ is a local martingale and
hence a martingale because it is bounded.
\end{pf}

\begin{defi}\label{d:localizing}
We call the probability measure $P^\star\sim P$ with density process
$Z^{P^\star}$ \textit{oppor\-tunity-neutral probability measure}.
\end{defi}

The opportunity-neutral probability measure is typically not a
martingale measure. In some instances it actually equals $P$ (cf.
Section \ref{su:ppstar}). For later use we determine the
$P^\star$-characteristics of $S$.
\begin{lemma}\label{l:SunterPstern}
The components of $S$ are locally $P^\star$-square integrable semimartingales.
Moreover,
\begin{eqnarray}
\label{e:bdach} &\displaystyle{b^{S\star} ={\bar b\over1+\Delta A^K},}&
\\
\label{e:cdach}&\displaystyle{\tilde c^{S\star}={\bar c\over1+\Delta
A^K},}&
\\
\label{e:adach} &\displaystyle{\bigl(1+(b^{S\star})^\top(\hat
c^{S\star})^{-1}b^{S\star}\Delta A \bigr)
 \bigl(1-(b^{S\star})^\top(\tilde c^{S\star})^{-1}b^{S\star}\Delta
A \bigr) =1}&
\end{eqnarray}
and
\begin{eqnarray}
\label{e:NA1} \hat c^{S\star}(\hat
c^{S\star})^{-1}b^{S\star}&=&b^{S\star },
\\
\label {e:NA2} \tilde c^{S\star}(\tilde c^{S\star})^{-1}b^{S\star}&=&b^{S\star },
\\
\label{e:NA3} \bar c\bar c^{-1}\bar b&=&\bar b.
\end{eqnarray}
$P\otimes A$-almost everywhere, where
\begin{eqnarray}\label{e:querb}
\bar b&:=&b^S+c^{SL}{1\over L_{-}} +\int x{y\over
L_{-}}F^{S,L}\bigl(d(x,y)\bigr)
\\
\label{e:querb2} &\hspace*{3pt} = &b^S+c^{SK}+\int
xyF^{S,K}\bigl(d(x,y)\bigr)
\end{eqnarray}
and
\begin{eqnarray}\label{e:querc}
\bar c&:=&c^{S}+\int xx^\top \biggl({1+{y\over L_{-}}}
\biggr)F^{S,L}\bigl(d(x,y)\bigr)
\\
\label{e:querc2} &\hspace*{3pt} =&c^{S}+\int
xx^\top(1+y)F^{S,K}\bigl(d(x,y)\bigr).
\end{eqnarray}
\end{lemma}

\begin{pf}The components of $S$ are locally
$P^\star$-square-integrable semimartingales because ${dP^\star\over
dP}=Z^{P^\star}_T$ is bounded (cf. Lemma~\ref{l:s2}). Let
\[
M:={1\over Z^{P^\star}_-}\mal Z^{P^\star}={1\over1+\Delta A^K}\mal M^K.
\]
Observe that
\[
K^c-M^c=K^c-{1\over1+\Delta A^K}\mal K^c={\Delta A^K\over1+\Delta
A^K}\mal K^c.
\]
Since
\begin{eqnarray*}
\bigg\langle{\Delta A^K\over1+\Delta A^K}\mal K^c, {\Delta
A^K\over1+\Delta A^K}\mal K^c \bigg\rangle_T &=& \biggl({\Delta
A^K\over1+\Delta A^K} \biggr)^2\mal\langle K^c,K^c\rangle_T
\\
&=&\sum_{t\le T} \biggl({\Delta A^K_t\over1+\Delta A^K_t} \biggr)^2
\Delta\langle K^c,K^c\rangle_t
\\  &=&0
\end{eqnarray*}
by continuity of $K^c$, we have ${\Delta A^K\over1+\Delta
A^K}\mal K^c=0$ and hence $M^c=K^c$. Moreover, $M$~is a local
martingale with $\Delta M={1\over1+\Delta A^K}\Delta K-{\Delta
A^K\over1+\Delta A^K}$. Together, it follows that
$b^{S,M}=(b^S,0)^\top$, $c^{S,M}=c^{S,K}$,
\[
F^{S,M}(G)= \int1_G \biggl(x,{y-\Delta A^K\over1+\Delta A^K}
\biggr)F^{S,K}\bigl(d(x,y)\bigr)
\]
for $G\in\BBB^{d+1}$ with $G\cap(\{0\}^d\times\rr)=\varnothing$. By the
Girsanov theorem as in Lemma~\ref{l:girsanov},
$P^\star$-characteristics $(b^{S\star},c^{S\star},F^{S\star},A)$ of $S$
are given by
\begin{eqnarray*}
b^{S\star}&=&b^S+c^{SM}+\int xyF^{S,M}\bigl(d(x,y)\bigr)
\\[-1pt]
&=&b^S+c^{SK}+\int x{y-\Delta A^K\over1+\Delta A^K}F^{S,K}\bigl(d(x,y)\bigr)
\\[-1pt]
&=&{1\over1+\Delta A^K}
 \biggl(b^S+c^{SK}+\int xyF^{S,K}\bigl(d(x,y)\bigr) \biggr),
\end{eqnarray*}
$c^{S\star}=c^S$ and
\begin{eqnarray*}
F^{S\star}(G)&=&\int1_G(x)(1+y)F^{S,M}\bigl(d(x,y)\bigr)
\\
&=&{1\over1+\Delta A^K}\int1_G(x)(1+y)F^{S,K}\bigl(d(x,y)\bigr)
\end{eqnarray*}
for $G\in\BBB^d$ with $0\notin G$. This yields (\ref{e:bdach}),
(\ref{e:cdach}).

Using the same argument as in the proof of
\cite{delbaenschachermayer95}, Theorem~3.5, it follows that
$b^{S\star}_t\in\hat c^{S\star}_t\rr^d$ and also $b^{S\star}_t\in\tilde
c^{S\star}_t\rr^d$ $(P\otimes A)$-almost everywhere on
$\Omega\times[0,T]$. (Due to Assumption \ref{a:eins} local boundedness
is not needed in our setup.) This implies (\ref{e:NA1}), (\ref{e:NA2}),
and hence also (\ref{e:NA3}) outside some $P\otimes A$-null set.
Consequently,
\begin{eqnarray*}
&& \bigl(1+(b^{S\star})^\top(\hat c^{S\star})^{-1}b^{S\star }\Delta A
\bigr) \bigl(1-(b^{S\star})^\top(\tilde
c^{S\star})^{-1}b^{S\star}\Delta A \bigr)
\\
&&\qquad =1+(b^{S\star})^\top \bigl((\hat c^{S\star})^{-1}-(\tilde
c^{S\star})^{-1}- (\hat
c^{S\star})^{-1}b^{S\star}(b^{S\star})^\top(\tilde c^{S\star
})^{-1}\Delta A \bigr) b^{S\star}\Delta A
\\
&&\qquad = 1+(b^{S\star})^\top (\hat c^{S\star})^{-1}
 \bigl(\tilde c^{S\star}-\hat c^{S\star}-b^{S\star}(b^{S\star
})^\top \Delta A \bigr) (\tilde c^{S\star})^{-1}b^{S\star}\Delta A
\\
&&\qquad = 1.
\end{eqnarray*}\upqed
\end{pf}

\begin{rem}\label{r:reicht}
An inspection of the proofs of Lemmas \ref{l:K} and
\ref{l:SunterPstern} yields that $L$ need not be the opportunity
process for (\ref{e:NA3}) to hold. We only used the fact that
$L=L_0\EEE(K)$ is a bounded semimartingale with $b^L\geq0$ and
$L,L_->0$.
\end{rem}

\subsection{\texorpdfstring{Characterization of $L$ and $\tilde a$.}{Characterization of $L$ and $\tilde
a$}} The opportunity process $L$ and the adjustment process $\tilde a$
play a crucial role in quadratic hedging. For example, they yield the
density processes of the variance-optimal S$\sigma
$MM 
$Q^\star$ and the opportunity-neutral
measure $P^\star$, which in turn lead to formulas for the optimal hedge
in Section \ref{s:hedge}.
The characterizations of $L$ and $\tilde a$ in this section help to determine
these processes in concrete models.
\begin{lemma}
We have
\begin{eqnarray}
\label{e:teil1} b^L&=&L_-\tilde a^\top\bar b,
\\
\label{e:teil2} \bar
b&=&\bar c\tilde a,
\\
\label{e:bK1} b^K&=&\bar b^\top\bar c^{-1}\bar b
=(b^{S\star})^\top(\hat c^{S\star})^{-1}b^{S\star}
\end{eqnarray}
outside some $P\otimes A$-null set, where $\bar b$, $\bar c$ are
defined in \textup{(\ref{e:querb})} and~\textup{(\ref{e:querc}).}
\end{lemma}

\begin{pf}
We denote by $\tau_n$ the stopping times in the proof of
Lemma~\ref{l:L}. Fix $n\in\nn$. Integration by parts and
Lemma~\ref{l:L} yield that
\[
\bigl(\EEE\bigl(\bigl(-\tilde a1_{\zu\tau_n,T\zu}\bigr)\mal
S\bigr)_-1_{\zu\tau_n,T\zu } \bigr)\mal \bigl(L-(L_-\tilde a)\mal
S-\tilde a\mal[L,S]\bigr) =1_{\zu\tau_n,T\zu}\mal M^{(\tau_n)}
\]
is a martingale.
Consequently, its compensator
\[
 \bigl(\EEE\bigl(\bigl(-\tilde a1_{\zu\tau_n,T\zu}\bigr)\mal S\bigr)_-
(b^L-L_-\tilde a^\top b^S-\tilde a^\top\tilde c^{SL})1_{\zu\tau _n,T\zu
} \bigr)\mal A
\]
vanishes. Since $\EEE((-\tilde a1_{\zu\tau_n,T\zu})\mal S)_-\not=0$ on
$\zu \tau _n,\tau _{n+1}\zu$, this implies that
\[
b^L-\tilde a^\top L_-b^S-\tilde a^\top\tilde c^{SL}=0
\]
$P\otimes A$-almost everywhere on $\zu\tau_n,\tau_{n+1}\zu$.
This yields (\ref{e:teil1}).

Fix $n\in\nn$. From Lemma~\ref{l:L}(2) and integration by parts it
follows that
\begin{eqnarray*}
0&\sim&1_{\zu\tau_n,T\zu}\mal\bigl(SM^{(\tau_n)}\bigr)
\\
&=&1_{\zu\tau_n,T\zu}\mal \bigl(S_-\mal M^{(\tau_n)}+M^{(\tau
_n)}_-\mal S+ \bigl[S,M^{(\tau_n)}\bigr] \bigr)
\\
&\sim&1_{\zu\tau_n,T\zu}\mal \bigl( \bigl(\EEE\bigl(\bigl(-a1_{\zu\tau
_n,T\zu}\bigr)\mal
S\bigr)_-L_- \bigr)\mal S + \bigl[S,\EEE\bigl(\bigl(-a1_{\zu\tau_n,T\zu}\bigr)\mal S\bigr)L \bigr] \bigr)
\\
&=& \bigl(\EEE\bigl(\bigl(-a1_{\zu\tau_n,T\zu}\bigr)\mal
S\bigr)_-1_{\zu\tau_n,T\zu } \bigr)
\\
&& {} \mal \bigl(L_-\mal S+[S,L]-\tilde
a\mal\bigl(L_-\mal[S,S]-\bigl[[L,S],S\bigr]\bigr) \bigr)
\\
&\sim& \biggl(\EEE\bigl(\bigl(-a1_{\zu\tau_n,T\zu}\bigr)\mal S\bigr)_-1_{\zu\tau_n,T\zu }
\\
&&\phantom{\biggl(} {}\times \biggl(L_-b^S+\tilde c^{SL}-
\biggl(L_-\tilde c^S+\int xx^\top yF^{S,L}\bigl(d(x,y)\bigr)
\biggr)\tilde a \biggr) \biggr)\mal A.
\end{eqnarray*}
Since $\EEE((-\tilde a1_{\zu\tau_n,T\zu})\mal S)_-$ does not vanish on
$\zu \tau_n,\tau_{n+1}\zu$, we have
\[
L_-b^S+\tilde c^{SL}- \biggl(L_-\tilde c^S+\int xx^\top
yF^{S,L}\bigl(d(x,y)\bigr) \biggr)\tilde a=0
\]
and hence (\ref{e:teil2}) outside some $P\otimes A$-null set.

Finally, (\ref{e:teil1}), (\ref{e:teil2}), (\ref{e:NA3}) yield
\[
L_-\bar b^\top\bar c^{-1}\bar b
=L_-\bar b^\top\bar c^{-1}\bar c \tilde a=L_-\bar b^\top\tilde a=b^L,
\]
which in turn implies the first equality in (\ref{e:bK1}).

On the set $\{\Delta A=0\}\supset\{\Delta A^K=0\}$, the second equality
follows from
(\ref{e:bdach}), (\ref{e:cdach}).
On $\{\Delta A^K\not=0\}$ the same equations yield
\[
1=(1+\Delta A^K)-b^K\Delta A =(1+\Delta
A^K)\bigl(1-(b^{S\star})^\top(\tilde c^{S\star})^{-1}b^{S\star }\Delta
A\bigr).
\]
In view of (\ref{e:adach}) we have
\[
1+b^K\Delta A=1+\Delta A^K=1+(b^{S\star})^\top(\hat c^{S\star
})^{-1}b^{S\star}\Delta A,
\]
which in turn implies $b^K=(b^{S\star})^\top(\hat c^{S\star
})^{-1}b^{S\star}$
on the set $\{\Delta A^K\not=0\}$.
\end{pf}

\begin{cor}\label{co:adjustment2}
The adjustment process and the extended adjustment process
satisfy the equations
\begin{equation}\label{e:teil3}
b^{S\star}=\tilde c^{S\star}\tilde a=\hat c^{S\star}\hat a
\end{equation}
or, put differently,
\[
\label{e:a}A^ {S\star}= \tilde a\mal\langle S,S \rangle^{P^\star}
= \hat a\mal\langle M^{S\star},M^{S\star}\rangle^{P^\star}.
\]
In the univariate case, this can be written more intuitively in terms
of pathwise Radon--Nikodym derivatives:
\[
\tilde a_t={dA^ {S\star}_t\over d\langle S,S
\rangle^{P^\star}_t},\qquad \hat a_t={dA^ {S\star}_t\over d\langle
M^{S\star},M^{S\star} \rangle ^{P^\star}_t}.
\]
\end{cor}

\begin{pf}$b^{S\star}=\tilde c^{S\star}\tilde a$ follows
from (\ref{e:teil2}), (\ref{e:bdach}), (\ref{e:cdach}). Together with
(\ref{e:NA2}), (\ref{e:adach}), (\ref{e:bK1}) we have
\begin{eqnarray*}
\hat c^{S\star}\tilde a &=& \bigl(\tilde
c^{S\star}-b^{S\star}(b^{S\star})^\top\Delta A \bigr)\tilde a
\\
&=&b^{S\star} \bigl(1-(b^{S\star})^\top(\tilde c^{S\star })^{-1}\tilde
c^{S\star}\tilde a\Delta A \bigr)
\\
&=&b^{S\star} \bigl(1-(b^{S\star})^\top(\tilde c^{S\star
})^{-1}b^{S\star }\Delta A \bigr)
\\
&=&b^{S\star}\over{1+(b^{S\star})^\top(\hat c^{S\star
})^{-1}b^{S\star }\Delta A}
\\
&=&b^{S\star}\over{1+\Delta A^K},
\end{eqnarray*}
which yields $b^{S\star}=\hat c^{S\star}\hat a$.
\end{pf}

\begin{lemma}\label{l:adach}
We have $\hat a\in L(M^{S\star})$.
\end{lemma}

\begin{pf}
Equations (\ref{e:teil3}), (\ref{e:bdach}), (\ref{e:teil1}) imply that
\[
(\hat a^\top\hat c^{S\star}\hat a)\mal A_T =\bigl((1+\Delta A^K)\tilde
a^\top b^{S\star}\bigr)\mal A_T =(\tilde a^\top\bar b)\mal A_T ={1\over
L_-}\mal A^L_T<\infty
\]
and hence $\hat a\in\lloc(M^{S\star})\subset L(M^{S\star})$ relative
to $P^\star$.
\end{pf}

\begin{defi}
We call
\[
N^\star:=-\hat a\mal M^{S\star}
\]
\textit{$P^\star$-minimal logarithm process}.
\end{defi}

The terminology is motivated by the fact that $\EEE(N^\star)$ is
essentially the density process of the so-called \textit{minimal signed
martingale measure} relative to $P^\star$ instead of $P$ (in the sense
of \cite{schweizer96}, (3.14)).

\begin{lemma}\label{l:VOMMMMM}
We have
\[
{L_0\over E(L_0)}\EEE(N)=Z^{P^\star}\EEE(N^\star).
\]
Consequently, $\EEE(N^\star)$ is the density process of $Q^\star$
relative to $P^\star$.
\end{lemma}

\begin{pf}
Integration by parts yields
\[
{L_0\EEE(N)\over E(L_0)Z^{P^\star}} ={\EEE(-\tilde a\mal
S)L\EEE(A^K)\over L} =\EEE (-\tilde a\mal S+A^K-[\tilde a\mal S,A^K] ).
\]
The term in parentheses on the right-hand side equals
\begin{eqnarray}\label{e:last}
&& x -\tilde a\mal M^{S\star}-(\tilde a^\top b^{S\star})\mal A
+b^K\mal A
\nonumber
\\[-8pt]
\\[-8pt]
\nonumber &&\qquad {}-(\tilde a\Delta A^K)\mal M^{S\star}-(\tilde
a^\top b^{S\star}\Delta A^K)\mal A
\end{eqnarray}
(cf. \cite{js87}, I.4.49b). Since
\[
b^K={1\over L_-}b^L=\tilde a^\top\bar b
=\tilde a^\top b^{S\star}(1+\Delta A^K)
\]
by (\ref{e:teil1}), (\ref{e:bdach}), the expression in (\ref{e:last})
equals $-\hat a\mal M^{S\star}=N^\star$.
\end{pf}

Roughly speaking, the next statement is another way of saying that
$S$ is a \mbox{$Q^\star$-$\sigma$-}martin\-gale.

\begin{lemma}\label{l:nstar}
$N^\star$ and $S+[S,N^\star]$ are \mbox{$P^\star$-$\sigma$-}martingales, which
implies that $S\EEE(N^\star)$ is a \mbox{$P^\star$-$\sigma $-}martingale as
well.
\end{lemma}

\begin{pf}$N^\star$ is a $P^\star$-$\sigma$-martingale by
definition. Moreover,
\begin{eqnarray*}
S+[S,N^\star] &=&S-\hat a\mal[S,M^{S\star}]
\\
&=&S-\hat a\mal[M^{S\star},M^{S\star}]- \hat a\mal \bigl((\Delta
A^{S\star})\mal M^{S\star} \bigr)
\\
&\hspace*{6pt} \sim^\star& (b^{S\star}-\hat
c^{S\star}\hat a ) \mal A=0
\end{eqnarray*}
by (\ref{e:teil3}). The last statement follows as in Lemma~\ref{l:N}.
\end{pf}

Corollary \ref{co:adjustment2} expresses the adjustment process in
terms of the $P^\star$-characteris\-tics of $S$. Of course this only
helps if the opportunity-neutral measure is known in the first place.
The following important result characterizes $L$ and $\tilde a$
directly in terms of $P$-characteristics.

\begin{teo}\label{t:unique}
The opportunity process is the unique semimartingale $L$ such that:
\begin{longlist}[3.]
\item[1.] $L,L_-$ are $(0,1]$-valued,

\item[2.] $L_T=1$,

\item[3.] The joint characteristics of $(S,L)$ solve the equation
\begin{equation}\label{e:L}
b^L=L_{-}\bar b^\top\bar c^{-1}\bar b
\end{equation}
outside some $P\otimes A$-null set, where $\bar b$, $\bar c$ are
defined as in \textup{(\ref{e:querb}), (\ref{e:querc}),}

\item[4.]
\begin{eqnarray}\label{e:int1}
& a\EEE\bigl(\bigl(-a1_{\zu\tau,T\zu}\bigr)\mal
S\bigr)_-1_{\zu\tau,T\zu}\in\overline \Theta ,&
\\
\label{e:int2} &\EEE\bigl(\bigl(-a1_{\zu\tau,T\zu}\bigr)\mal
S\bigr)L\mbox{ is of class (D)}&
\end{eqnarray}
hold for $a:=\bar c^{-1}\bar b$ and any stopping time $\tau$.
\end{longlist}
In this case $a=\bar c^{-1}\bar b$ meets the requirement of an
adjustment process $\tilde a$ in Lemma~\textup{\ref{l:exa}.}
\end{teo}

\begin{pf}Suppose that $L$ is the opportunity process.
Properties 1 and 2 are shown in Lemmas \ref{l:L} and \ref{l:values}.
Equation (\ref{e:bK1}) and $b^L=L_-b^K$ yield (\ref{e:L}).
By~(\ref{e:bdach}), (\ref{e:cdach}), (\ref{e:NA2}), (\ref{e:bK1}) we
have
\begin{eqnarray*}
(a^\top\hat c^{S\star}a)\mal A_T &\leq& (a^\top\tilde c^{S\star}a)\mal
A_T =  ((b^{S\star})^\top(\tilde c^{S\star})^{-1}b^{S\star} )\mal A_T
\\
&=& {1\over1+\Delta A^K}\mal A^K_T<\infty,
\end{eqnarray*}
which implies $a\in\lloc(M^{S\star})$ relative to $P^\star$ by
\cite{js87}, III.4.3. Similarly, we have $a\in L(A^{S\star})$ because
$|a^\top b^{S\star}|\mal A_T\leq{1\over1+\Delta A^K}\mal A^K_T<\infty$.
Together, it follows that $a\in L(S)$.

More specifically, we have
\[
a\mal A^{S\star}=(a^\top b^{S\star})\mal A
={b^L\over(1+\Delta A^K)L_-}\mal A
\]
and likewise for $\tilde a$ by (\ref{e:bdach}), (\ref{e:teil1}).
Similarly, (\ref{e:teil1}--\ref{e:bK1}) yield 
\[
 \langle(a-\tilde a)\mal M^{S\star}, (a-\tilde a)\mal M^{S\star
} \rangle^{P^\star} \leq \bigl((a-\tilde a)^\top\tilde
c^{S\star}(a-\tilde a) \bigr)\mal A=0,
\]
which implies $(a-\tilde a)\mal M^{S\star}=0$. Together, we have $a\mal
S=\tilde a\mal S$. Hence one may choose $\tilde a=a$ in
Lemma~\ref{l:exa}.

Finally, (\ref{e:int1}) follows from (\ref{e:mayassume}) and
(\ref{e:int2}) from Lemma~\ref{l:L}.

Conversely, let $L'$ be a semimartingale satisfying properties 1--4
with $\bar b',\bar c'$ as in~(\ref{e:querb}) and  (\ref{e:querc}).
Define $K':={1\over L'_-}\mal L'$ and $N':=K'-a\mal S-[a\mal S,K']$. We
use the notation $L', \bar b',\bar c', K', N'$ in this
part of the proof because $L'$ is yet to be shown to coincide with the
true opportunity process. From
\[
[S,K']={1\over L'_-}\mal[M^S,M^{L'}]+(\Delta A^S)\mal M^{K'}+(\Delta
A^{K'})\mal S
\]
and standard results (cf. \cite{js87}, I.4.24, III.3.14) it follows
that
\[
[S,K']=[S^{c},K'^c]+\int_{[0,\cdot]\times\rr^d\times\rr} xy\mu
^{(S,K')}\bigl(d(t,x,y)\bigr)
\]
is an $\rr^d$-valued special semimartingale with compensator
$(c^{SK'}+\int xyF^{S,K'}\times (d(x,y)))\mal A$.
For $n\in\nn$ define the predictable set $D_n:=\{|a|\leq n\}$.
Since $1_{D_n}$ and $a1_{D_n}$ are
bounded, we have that
\[
1_{D_n}\mal N'=1_{D_n}\mal K'-(1_{D_n}a)\mal S-(1_{D_n}a)\mal[S,K']
\]
is a special semimartingale as well with compensator
\begin{eqnarray*}
&& \biggl(1_{D_n}b^{K'}-1_{D_n}a^\top \biggl(b^S+c^{SK'}+\int
xyF^{S,K'}\bigl(d(x,y)\bigr) \biggr) \biggr)\mal A
\\
&&\qquad = \biggl( \biggl({b^{L'}\over L'_-}-\bar b'^\top\bar
c'^{-1}\bar b' \biggr)1_{D_n} \biggr)\mal A =0.
\end{eqnarray*}
Consequently, $1_{D_n}\mal N'$ is actually a local martingale. Since
$D_n\uparrow\Omega\times[0,T]$ up to an evanescent set, $N'$ is a
$\sigma$-martingale (cf. Remark~\ref{r:sigmamartingal}).

Similarly, we have that
\begin{eqnarray*}
&& 1_{D_n}\mal(S^i+[S^i,N'])
\\
&&\qquad = 1_{D_n}\mal S^i + 1_{D_n}\mal[S^i,K']
\\
&&\qquad\quad{} - \sum_{j=1}^n
(1_{D_n}a^j)\mal[S^i,S^j] - \sum_{j=1}^n (1_{D_n}a^j)\mal
\bigl[S^i,[S^j,K']\bigr]
\end{eqnarray*}
is a special semimartingale with compensator
\begin{eqnarray*}
&& \biggl( 1_{D_n}\biggl( b^S+c^{SK'}+\int xyF^{S,K'}\bigl(d(x,y)\bigr)
\\
&&\hspace*{9.3mm} {}
-c^Sa-\int x(x^\top a)(1+y)F^{S,K'}\bigl(d(x,y)\bigr) \biggr)^i \biggr)\mal A
\\
&&\qquad = \bigl(1_{D_n}(\bar b'-\bar c'a)^i \bigr)\mal A
\end{eqnarray*}
for $i=1,\dots,d$. Since $\bar b'-\bar c'a=\bar b'-\bar c'\bar
c'^{-1}\bar b'=0$ by Remark~\ref{r:reicht},
it follows that the process 
$1_{D_n}\mal(S^i+[S^i,N'])$
is a local martingale. This implies that $S+[S,N']$ is a $\sigma
$-martingale as well.

Fix a stopping time $\tau$. Let
$\vartheta:=a\EEE(-a1_{\zu\tau,T\zu}\mal S)_-1_{\zu\tau,T\zu}$ and
\[
Z:=(1-\vartheta\mal S)L'=\EEE\bigl(\bigl(-a1_{\zu\tau,T\zu}\bigr)\mal
S\bigr)L'.
\]
[In (\ref{e:int1}) and (\ref{e:int2}) it is implicitly assumed that
$a\in L(S)$ for the integral to make sense. By similar arguments as in
the first part of the proof one can show that this integrability
condition is in fact implied by properties 1--3 of
Theorem~\ref{t:unique}.]

Since $N'$ and $S+[S,N']$ are $\sigma$-martingales,
\[
{Z\over Z^\tau}=\EEE\bigl(1_{\zu\tau,T\zu}\mal
K'\bigr)\EEE\bigl(\bigl(-a1_{\zu\tau ,T\zu}\bigr)\mal
S\bigr)=\EEE(N'-N'^\tau)
\]
and
\begin{eqnarray*}
&& {Z\over Z^\tau}(S-S^\tau)
\\
&&\qquad  =\EEE(N'-N'^\tau)_-\mal
\bigl((S-S^\tau)_-\mal(N'-N'^\tau)+1_{\zu \tau,T\zu }\mal(S+[S,N'])
\bigr)
\end{eqnarray*}
are $\sigma$-martingales as well.

We show that $\vartheta$ is efficient on $\zu\tau,T\zu$. Indeed, from
(\ref{e:int2}) and Lemma~\ref{l:martingale} it follows that
$Z-Z^\tau=(Z_\tau1_{\zu\tau,T\zu})\mal{Z\over Z^\tau}$ is a martingale.
It is even a square-integrable martingale because $Z_T-Z_\tau\in
L^2(P)$. Let $\psi$ be a simple strategy with
$\psi1_{\auf0,\tau\zu}=0$. The same arguments as in step 1 of the proof
of Lemma~\ref{l:k20} yield that $(\psi\mal
S)Z=((Z_\tau\psi)\mal(S-S^\tau)){Z\over Z^\tau}$ is a martingale.
Consequently,
\begin{eqnarray*}
&& E \bigl(\bigl(1-(\vartheta+\psi)\mal S_T\bigr)^2 \bigr)
\\
&&\qquad \geq E \bigl((1-\vartheta\mal S_T)^2 \bigr)- 2E
\bigl((1-\vartheta\mal S_T)L'_T(\psi\mal S_T) \bigr)
\\
&&\qquad = E \bigl((1-\vartheta\mal S_T)^2 \bigr),
\end{eqnarray*}
which implies the optimality of $\vartheta$. Since $Z-Z^\tau$ is a
martingale, Lemma~\ref{l:L} yields that $L'$ is the opportunity
process.
\end{pf}

Condition (\ref{e:int1}) looks somewhat unpleasant because of the
involved definition of $\overline\Theta$. The following example shows
that uniqueness in Theorem~\ref{t:unique} does not generally hold
without this condition. For related considerations see also
\cite{schweizer96}~and~\cite{cernykallsen06bwp}.

\begin{exa}\label{ex:L}
Let $T=1$ and $S$ be a standard Wiener process. By
Theorem~\ref{t:unique} the opportunity and adjustment processes are
$L=1$ and $\tilde a=0$. Choose some doubling-type strategy $\psi\in
L(S)$ with $1-\psi\mal S\geq{1\over2}$ and $1-\psi\mal S_T={1\over2}$.
Of course, $\psi$ cannot be admissible. We write $1-\psi\mal
S=\EEE(-\bar a\mal S)$ with $\bar a:={\psi\over 1-\psi\mal S_-}$.
Define
\[
\overline L:={1\over2\EEE(-\bar a\mal S)}={1\over2}\EEE (\bar a\mal S+\bar
a^2\mal[S,S] ).
\]
Straightforward calculations yield that $\overline L$ satisfies
conditions 1--3 in Theorem~\ref{t:unique}. Moreover, $\bar a$ is the
corresponding process in condition 4. Since $\EEE((-\bar
a1_{\zu\tau,T\zu})\mal S)\overline L=\overline L^\tau$ is bounded,
(\ref{e:int2}) is satisfied as well.

It is interesting to note that the ``variance-optimal logarithm
process'' $\overline N$ corresponding to this wrong choice of $\overline
L,\bar a$ satisfies $\EEE(\overline N)={\overline L\over\overline
L_0}\EEE(-\bar a\mal S)=1$, that is, it coincides with the true
variance-optimal logarithm process. In particular,
\[
{dQ^\star\over dP}
={1-\psi\mal S_T\over E(1-\psi\mal S_T)},
\]
which parallels the last expression in (\ref{e:VOMM}).
Nevertheless, $\psi$ is not an efficient strategy on $\zu0,T\zu$
because it is not admissible.
\end{exa}

In concrete models, it may be easier to verify the following sufficient
condition
instead of (\ref{e:int1}), (\ref{e:int2}).

\begin{lemma}\label{l:sufficient}
Let $L$ be a special semimartingale satisfying conditions \textup{1--3}
in Theorem~\textup{\ref{t:unique}} with $\bar b,\bar c$ defined as in
\textup{(\ref{e:querb}), (\ref{e:querc})}. If $a:=\bar c^{-1}\bar b$
satisfies
\[
\sup \bigl\{E \bigl(\EEE\bigl(\bigl(-a1_{\zu\tau,T\zu}\bigr)\mal
S\bigr)_\sigma ^2 \bigr)\dvtx  \sigma\mbox{ stopping time}
\bigr\}<\infty
\]
for any stopping time $\tau$, then condition \textup{4} holds as well,
that is, $L$ is the opportunity process.
\end{lemma}
\begin{pf}Condition (\ref{e:int2}) is obvious because $L$
is bounded. Let $Q$ be an S$\sigma$MM with density process $Z^Q$ and
${dQ\over dP}\in L^2(P)$. Integration by parts yields that
$(\vartheta\mal S)Z^Q$ is a $\sigma$-martingale for
\[
\vartheta:= a\EEE \bigl(\bigl(- a1_{\zu\tau,T\zu}\bigr)\mal S
\bigr)_-1_{\zu \tau,T\zu}
\]
[cf. (\ref{e:partial})]. Since $\sup_{t\in[0,T]}|Z^Q_t|\in L^2(P)$ by
Doob's inequality and $1-\vartheta\mal S$ is an $L^2$-semimartingale,
we have that $(\vartheta\mal S)Z^Q$ is of class (D) and hence a
martingale (cf. Lemma~\ref{l:martingale}). Using Corollary
\ref{co:theta} we obtain (\ref{e:int1}).
\end{pf}

\subsection{\texorpdfstring{When does $P^\star=P$ hold\textup{?}}{When does $P^\star=P$ hold}}\label{su:ppstar}
The opportunity-neutral measure plays a key role in quadratic hedging.
Therefore we want to have a closer look at the question when $P^\star$
equals $P$. In line with \cite{schweizer94}, we call
\[
\widehat K:= ((b^S)^\top(\hat c^S)^{-1}b^S )\mal A
\]
\textit{mean-variance tradeoff \textup{(}MVT\textup{)} process}.
Similarly, the MVT process relative to $P^\star$ is denoted by
$\widehat K^\star$, that is,
\[
\widehat K^\star:= ((b^{S\star})^\top(\hat c^{S\star })^{-1}b^{S\star }
)\mal A.
\]
Observe that $\widehat K^\star=A^K$ by (\ref{e:bK1}).

\begin{prop}\label{p:PP*}
The following statements are equivalent:
\begin{longlist}[6.]
\item[1.] $P^\star=P$.

\item[2.] $K$ (or equivalently $L$) is a predictable process of finite
variation and $L_0$ is deterministic.

\item[3.] $K=\widehat K$ and $L_0$ is deterministic.

\item[4.] $K=\widehat K^\star$ and $L_0$ is deterministic.

\item[5.] $\EEE(\widehat K)_T$ is finite and deterministic.

\item[6.] $\EEE(\widehat K^\star)_T$ is deterministic.
\end{longlist}
In this case the opportunity process equals $L={\EEE(\widehat K)/\EEE(\widehat K)_T}$.
\end{prop}

\begin{pf}
1${}\Rightarrow{}$2: Since $1=Z^{P\star}={L/(E(L_0)\EEE(A^K))}$, we
have that $L$ and hence also $K=\LLL(L)$ are predictable processes of
finite variation. $L_0$ is deterministic because $Z^{P\star}_0=1$.
{\smallskipamount=0pt
\begin{longlist}[2${}\Rightarrow{}$4:]
\item[2${}\Rightarrow{}$4:] This is obvious because $K=A^K=\widehat
K^\star$.

\item[4${}\Rightarrow{}$1:] This follows from
\[
Z^{P\star}={L\over E(L_0)\EEE(A^K)} ={L_0\EEE(K)\over E(L_0)\EEE(\widehat K^\star)}=1.
\]

\item[1${}\Rightarrow{}$6:] This follows from $Z^{P^\star}_T=1$ and
$\widehat K^\star=A^K$.

\item[6${}\Rightarrow{}$1:] This holds because $Z^{P\star}_T={1/
(E(L_0)\EEE(\widehat K^\star)_T)}$ is deterministic.

\item[1${}\Rightarrow{}$3:] In view of (1${}\Rightarrow{}$2), this
follows from $K=A^K=\widehat K^\star =\widehat K$.

\item[3${}\Rightarrow{}$5:] This follows from $1=L_T=L_0\EEE(K)_T$.

\item[5${}\Rightarrow{}$2:] Let $L:={\EEE(\widehat K)/\EEE(\widehat
K)_T}$. Since $\widehat K$ is an increasing predictable process, $L$ is
a $(0,1]$-valued increasing predictable process. The predictability of
$L$ implies $c^{SL}=0$ and $y=\Delta L_t$
$(F_t^{S,L}(d(x,y))A(dt))$-almost everywhere. If $\bar b$, $\bar c$ are
defined as in (\ref{e:querb}), (\ref{e:querc}), we have $\bar
b=(1+\Delta\widehat K)b^S$, $\bar c=(1+\Delta\widehat K)\tilde c^S$ and
hence
\[
L_-\bar b^\top\bar c^{-1}\bar b= L_- \bigl(1+(b^S)^\top(\hat
c^S)^{-1}b^S\Delta A \bigr)(b^S)^\top (\tilde c^S)^{-1}b^S.
\]
\end{longlist}}
Observe that (\ref{e:adach}--\ref{e:NA2}) can be derived literally
for $P$ instead of $P^\star$. We obtain
\[
L_-\bar b^\top\bar c^{-1}\bar b=L_-(b^S)^\top(\hat c^S)^{-1}b^S=b^L,
\]
which implies that $L$ satisfies conditions 1--3 in
Theorem~\ref{t:unique}. If we can show that $L$ is the true opportunity
process, then $P^\star=P$ follows from Lemma~\ref{l:K}.

Fix any stopping time $\tau$.
For $a:=\bar c^{-1}\bar b=(\tilde c^{S})^{-1}b^{S\star}$ and
$X:=(-a1_{\zu\tau,T\zu})\mal S$ we have
\begin{eqnarray*}
\langle M^X,M^X\rangle_T&=& \bigl(a^\top\hat c^Sa1_{\zu\tau,T\zu}
\bigr)\mal A_T
\\
&\leq& \bigl(a^\top\tilde c^Sa1_{\zu\tau,T\zu} \bigr)\mal A_T
\\
&=& \bigl((b^S)^\top(\tilde c^S)^{-1}\tilde c^S(\tilde
c^S)^{-1}b^S1_{\zu \tau ,T\zu} \bigr)\mal A_T
\\
&=& \biggl({1_{\zu\tau,T\zu}\over1+\Delta\widehat K}\bar b^\top\bar
c^{-1}\bar b \biggr)\mal A_T
\\
&\leq& ((b^S)^\top(\hat c^S)^{-1}b^S )\mal A_T
\\
&=&\widehat K_T\leq\EEE(\widehat K)_T.
\end{eqnarray*}
Similarly, we have
\[
\operatorname{var}(A^X)_T = \big|a^\top b^S1_{\zu\tau,T\zu} \big|\mal
A_T = \biggl({1_{\zu\tau,T\zu}\over1+\Delta\widehat K}\bar b^\top\bar
c^{-1}\bar b \biggr)\mal A_T \leq\EEE(\widehat K)_T
\]
for the variation process of $A^X$. In view of Lemmas \ref{l:jacod} and
\ref{l:sufficient}, $L$ is the opportunity process.
\end{pf}

To relate the condition $P^\star=P$ to earlier literature, we define
(myopic) portfolio weights
\begin{eqnarray}\label{e:modlambda}
\tilde\lambda&:=&(\tilde c^S)^{-1}b^S,
\nonumber
\\[-8pt]
\\[-8pt]
\nonumber
\hat\lambda&:=&(1+\Delta\widehat K)\tilde\lambda
\end{eqnarray}
in accordance with \cite{schweizer94}. Repeating the arguments leading
to (\ref{e:teil3}) under $P$ rather than $P^\star$ yields $\hat
c^S\hat\lambda=b^S$ [which implies that $\hat\lambda=(\hat
c^S)^{-1}b^S$ if $\hat c^S$ is invertible]. By Theorem~1 of
\cite{schweizer95b} we have $\hat\lambda\in L(M^S)$.

\begin{defi}
If $\EEE(-\hat\lambda\mal M^S)$ is of class (D) and hence a martingale,
then it is the density process of some S$\sigma$MM $Q$. Only slightly
extending \cite{schweizer96}, (3.14) we call $Q$ the \textit{minimal
signed martingale measure \textup{(}minimal S$\sigma$MM\textup{)}}.
\end{defi}

In view of Proposition~\ref{p:PP*}, the following corollary can be
interpreted as an extension of Proposition~5.1 in \cite{laurentpham99}.
It also extends sufficient conditions for $Q^\star=Q$ given in
\cite{schweizer96}, Examples 1 and 2.

\begin{cor}\label{co:MMM}
Suppose $\EEE(-\tilde a\mal S)_T\not=0$ almost surely. Then there is
equivalence between:
\begin{longlist}[2.]
\item[1.] $P^\star=P$,

\item[2.] $\widehat K_T$ is finite, the minimal S$\sigma$MM 
$Q$ exists, $Q^\star=Q$, and $\tilde a$ can be chosen as~$\tilde
\lambda$.
\end{longlist}
The implication \textup{1}${}\Rightarrow{}$\textup{2} still holds
without the assumption on $\EEE(-\tilde a\mal S)$.
\end{cor}

\begin{pf}
1${}\Rightarrow{}$2: This follows from Lemma~\ref{l:VOMMMMM},
Theorem~\ref{t:unique}, and (\ref{e:bdach}), (\ref{e:cdach}).

2${}\Rightarrow{}$1: As in the proof of Lemma~\ref{l:SunterPstern}
it follows that $b^S=\hat c^S(\hat c^S)^{-1}b^S$ and hence
$\hat\lambda^\top b^S=(b^S)^\top(\hat c^S)^{-1}b^S$. Hence, the density
process of $Q$ equals
\begin{eqnarray*}
\EEE(-\hat\lambda\mal M^S)&=& \EEE \bigl((\hat\lambda^\top b^S)\mal
A-\hat\lambda\mal S \bigr)
\\
&=&\EEE \bigl(((b^S)^\top(\hat c^S)^{-1}b^S)\mal
A-\bigl((1+\Delta\widehat K)\tilde\lambda\bigr)\mal S \bigr)
\\
&=&\EEE \bigl(\widehat K-\tilde\lambda\mal S-(\Delta\widehat K)\mal(\tilde \lambda \mal S) \bigr)
\\
&=&\EEE (\widehat K-\tilde\lambda\mal S-[\widehat K,\tilde\lambda\mal S] )
\\
&=&\EEE(\widehat K)\EEE(-\tilde\lambda\mal S),
\end{eqnarray*}
where the fourth equality follows from \cite{js87}, I.4.49b and the
last from Yor's formula. This density process equals ${L\over
E(L_0)}\EEE(-\tilde a\mal S)$ by $Q^\star=Q$ and
Proposition~\ref{p:VOMM}. Since $\tilde a=\tilde\lambda$ and
$\EEE(-\tilde a\mal S)$ never vanishes (cf. \cite{js87}, I.4.61), we
have that $L=E(L_0)\EEE(\widehat K)$ is predictable with $L_0=E(L_0)$.
The assertion follows now from Proposition~\ref{p:PP*}
(2${}\Rightarrow{}$1).
\end{pf}

Finally, we consider the situation of deterministic mean-variance
tradeoff, which is the focus of \cite{schweizer94}.
\begin{cor}\label{co:DMVT}
If the MVT process $\widehat K$ is finite and deterministic, then
$L:={\EEE(\widehat K)/\EEE(\widehat K)_T}$ is the opportunity  process,
$K:=\widehat K$ is the modified mean-variance tradeoff process, and
$P^\star=P$.
\end{cor}
\begin{pf}This follows from Proposition~\ref{p:PP*}
(5${}\Rightarrow{}$1, 3) and from $1=L_T=L_0\EEE(K)_T$. \end{pf}

\subsection{\texorpdfstring{Determination of the opportunity
process.}{Determination of the opportunity process}}\label{su:L} Unless
we are in the fortunate situation of Corollary \ref{co:DMVT} or at
least Proposition~\ref{p:PP*}, the crucial step in concrete
applications is to determine the opportunity process $L$. This is
relatively easy in discrete time.

\begin{exa}
Suppose that we are actually considering a discrete-time model, that
is, $A_t=[t]:=\max\{n\in\nn\dvtx n\leq t\}$ and $\FFF_t=\FFF_{[t]}$ for
$t\in[0,T]$ with $T\in\nn$. In this case all processes in this paper
are (or can be chosen) piecewise constant between integer times. For
ease of notation suppose that $d=1$ (only one tradable asset). By
\cite{js87}, II.3.11 we have $b^L_t=E(\Delta L_t|\FFF_{t-1})$,
$\bar b_t=E(\Delta S_t L_t/L_{t-1}|\FFF_{t-1})$, and 
$\bar c_t=E((\Delta S_t)^2 L_t/L_{t-1}|\FFF_{t-1})$ for
$t\in\{1,2,\dots,T\}$. Consequently, (\ref{e:L}) can be rewritten as
\begin{equation}\label{e:recursive}
L_{t-1}=E(L_t|\FFF_{t-1})-{ (E(\Delta S_t L_t|\FFF_{t-1}) )^2\over
E((\Delta S_t)^2 L_t|\FFF_{t-1})},
\end{equation}
that is, the opportunity process is determined by a simple backward
recursion starting in $L_T=1$. For the adjustment process we have
\[
\tilde a_t={\bar b_t\over\bar c_t}={E(\Delta S_t L_t|\FFF_{t-1})\over
E((\Delta S_t)^2 L_t|\FFF_{t-1})}.
\]
\end{exa}

The previous example indicates that the characteristic equation
(\ref{e:L}) may be interpreted as the continuous-time analogue of a
backward recursion. True continuous-time models are typically Markovian
in $S_t$ or at least $(S_t,Y_t)$ with some additional process $Y$ as,
for example, stochastic volatility. If one makes the natural assumption
$L_t=f(t,S_t,Y_t)$ with some $C^2$-function $f$, then (\ref{e:L}) can
be rewritten as an integro-differential equation for $f$ by means of
It\^o's formula. But as it is not obvious whether the smoothness
assumption is justified, it may require substantial effort to make this
statement precise. In \cite{cernykallsen06awp} and ongoing research,
$L$ is determined explicitly by an ansatz of the above type in specific
stochastic volatility models.

Alternatively, the process $L$ can be interpreted as the solution to
some backward stochastic differential equation (BSDE). To this end, we
use the martingale representation theorem (cf. \cite{js87}, III.4.24)
to write the martingale part of $L$ as
\[
M^L=J\mal S^c+W*(\mu^S-\nu^S)+U
\]
with some $J\in\lloc(S^c)$, $W\in\gl(\mu^S)$ and some local martingale
$U\in\hloc$ such that $\langle U^c,S^c\rangle=0$ and
$M^P_{\mu^S}(\Delta U|\widetilde\PPP )=0$ in the sense of \cite{js87},
III.3c. Using the notation
\[
\widehat W_t:=E\bigl(W(t,\Delta S_t)|\FFF_{t-}\bigr),
\]
the quadruple $(J,W,L,U)$ solves the BSDE
\begin{eqnarray}\label{e:bsde}
\nonumber L&=&J\mal S^c+W*(\mu^S-\nu^S)+U
\\
\nonumber &&{}+ \biggl( \biggl(b^S+c^S {J\over L_-} +\int
{W(x)-\widehat W\over L_-}xF^S(dx) \biggr)^\top
\\
&&\phantom{{}+ \biggl(}
{}\times \biggl(c^S+\int xx^\top\biggl(1+
{W(x)-\widehat W\over L_-}\biggr)F^S(dx) \biggr)^{-1}
\\
\nonumber &&\phantom{{}+ \biggl(} {}\times \biggl(b^S+c^S {J\over L_-}
+\int {W(x)-\widehat W\over L_-}xF^S(dx) \biggr)L_- \biggr)\mal A,
\\
\nonumber L_T&=&1.
\end{eqnarray}
However, it is not obvious whether this representation is of any use.

One should note that (\ref{e:bsde}) is not related to the BSDEs (3.6)
and (4.10) in \cite{schweizer96}, which characterize the adjustment
process and the optimal hedge. The latter are hard to use in practice
because their terminal values involve the $L^2$-projection of~$1$,
respectively, $H$ on $K_2(0)$, which is generally unknown. If at all,
one may rather observe a certain similarity between (\ref{e:recursive})
and the recursive expression (2.1) in \cite{schweizer96} for the
adjustment process in discrete time. Mania and Tevzadze
\cite{maniatevzadze00,maniatevzadze03a} derive BSDE's for $1/L$ in the
case of a continuous asset price process $S$. These equations are quite
different from both (\ref{e:bsde}) and (\ref{e:L}).

\section{\texorpdfstring{On the quadratic hedging problem.}{On the quadratic hedging problem}} \label{s:hedge}
We now come back to the hedging problem from Definition~\ref{d:hedge}.
The processes and measures $\lambda^{(\tau)}$, $M^{(\tau)}$, $L$,
$\tilde a$, $\hat a$, $K$, $N$, $Q^\star$, $Z^{P^\star}$, $P^\star$,
$\bar b$, $\bar c$ are defined as in the previous section. Recall that
$P$ is the default probability measure for expectations, martingales
and so forth.

\subsection{\texorpdfstring{Mean value process and pure hedge
coefficient.}{Mean value process and pure hedge coefficient}} If $S$ is
a martingale, the mean value process $V_t=E(H|\FFF_t)$ leads to the
optimal hedge via (\ref{e:kunitawatanabe}). If $S$ fails to be a
martingale, a similar role is played by the conditional expectation
of $H$ relative to the variance-optimal S$\sigma$MM 
$Q^\star$. By the generalized Bayes formula we have
\begin{equation}\label{e:condexp}
E_{Q^\star}(H|\FFF_t)=E \bigl( H\EEE(N-N^t)_T |\FFF_t \bigr)
\end{equation}
if $Q^\star$ is a true probability measure.
In the general case we use the right-hand side of (\ref{e:condexp})
as a substitute for the possibly undefined conditional expectation.

\begin{lemma}\label{l:V}
There is a unique semimartingale $V$ satisfying
\begin{eqnarray}\label{e:V}
V_t&=&E \bigl( H\EEE(N-N^t)_T |\FFF_t \bigr)
\\
\label{e:V2} &=&E_{P^\star} \bigl(
H\EEE\bigl(N^\star-(N^\star)^t\bigr)_T |\FFF _t \bigr)
\end{eqnarray}
for $t\in[0,T]$.
Moreover, $(V_sM^{(t)}_s)_{s\in[t,T]}$ is a martingale for any $t\in[0,T]$.
\end{lemma}

\begin{pf}
In this proof $\varphi$ denotes an optimal
hedging strategy for arbitrary initial endowment $v_0\in
L^2(\Omega,\FFF_0,P)$ or, alternatively, $(v_0,\varphi)$ denotes an
optimal endowment/strategy pair. Moreover, let $G:=v_0+\varphi\mal S$
and define a square-integrable martingale $Z$ by its terminal value
$Z_T:=G_T-H$. Finally, we set $V:=G-{Z\over L}$. The optimality of
$\varphi$ implies that
\begin{eqnarray*}
0&\leq&E \bigl((G_T+\varepsilon\vartheta\mal S_T-H)^2 \bigr)-E
\bigl((G_T-H)^2 \bigr)
\\
&=& 2\varepsilon E \bigl((\vartheta\mal S_T)Z_T \bigr) +\varepsilon^2 E
\bigl((\vartheta\mal S_T)^2 \bigr)
\end{eqnarray*}
for any simple strategy $\vartheta$ and any $\varepsilon\in\rr$.
Therefore
\begin{equation}\label{e:optimality}
E \bigl((\vartheta\mal S_T)Z_T \bigr)=0
\end{equation}
for any simple $\vartheta$, which implies that $SZ$ is a
$\sigma$-martingale. By Remark~\ref{r:erweiterung} we have that
$(\vartheta\mal S)Z$ is a martingale for any
$\vartheta\in\overline\Theta$. In particular,
$(G-V)M^{(t)}=(1-\lambda^{(t)}\mal S)Z$ is a martingale for any fixed
$t\in[0,T]$. By Lemma~\ref{l:L}(3), $(G_sM^{(t)}_s))_{s\in[t,T]}$ and
hence also $(V_sM^{(t)}_s))_{s\in[t,T]}$ is a martingale. Using
Lemma~\ref{l:exa}, we have
\begin{eqnarray*}
E \bigl( H\EEE(N-N^t)_T |\FFF_t \bigr)L_t &=&E \bigl(
V_TL_T\bigl(1-\lambda^{(t)}\mal S_T\bigr) |\FFF_t \bigr)
\\
&=&V_tL_t\bigl(1-\lambda^{(t)}\mal S_t\bigr)
\\
&=&V_tL_t,
\end{eqnarray*}
which shows (\ref{e:V}).

Along the same lines as Lemma~\ref{l:VOMMMMM} it follows that
\begin{equation}\label{e:samelines}
\EEE(N-N^t)=\EEE \bigl(N^\star-(N^\star)^t \bigr){Z^{P^\star}\over
(Z^{P^\star})^t}.
\end{equation}
Consequently,
\begin{eqnarray*}
E_{P^\star} \bigl( H\EEE\bigl(N^\star-(N^\star)^t\bigr)_T |\FFF _t
\bigr) &=&E \biggl( H\EEE \bigl(N^\star-(N^\star)^t
\bigr)_T{Z^{P^\star}_T \over Z^{P^\star}_t} \Big|\FFF_t \biggr)
\\
&=&E \bigl( H\EEE(N-N^t)_T |\FFF_t \bigr),
\end{eqnarray*}
which yields (\ref{e:V2}). The uniqueness (up to indistinguishability)
of $V$ is obvious.
\end{pf}

\begin{defi}\label{d:V}
We call $V$ from Lemma~\ref{l:V} \textit{mean value process} of the
option.
\end{defi}

The following technical statements mean essentially that $V$ is a
\mbox{$Q^\star$-$\sigma$-}mar\-tingale.

\begin{lemma}\label{l:VN}
We have \textup{1.} $V+[V,N]$ and hence $V\EEE(N)$ are $\sigma$-martingales.

\textup{2.} $V+[V,N^\star]$ and hence $V\EEE(N^\star)$ are $P^\star $-$\sigma
$-martingales.
\end{lemma}

\begin{pf}%
1. Fix $n\in\nn$. If $(\tau_n)_{n\in\nn}$ denotes the sequence of
stopping times from the proof of Lemma~\ref{l:L}, then
\[
\EEE(N-N^{\tau_n})_-={L_-\over L^{\tau_n}_-}\bigl(1-\lambda^{(\tau
_n)}\mal S_-\bigr)\not=0
\]
on $\zu\tau_n,\tau_{n+1}\zu$.
For $t\in\zu\tau_n,T\zu$ we have
\[
\EEE(N-N^t)\EEE(N-N^{\tau_n})_t=\EEE(N-N^{\tau_n}),
\]
which implies that
$(L_{\tau_n}1_{\zu\tau_n,T\zu})\mal(V\EEE(N-N^{\tau_n}))$ is a
martingale by (\ref{e:V}). Integration by parts and Lemma~\ref{l:N}
yield that
\begin{eqnarray*}
&& 1_{\zu\tau_n,\tau_{n+1}\zu}\mal(V+[V,N])
\\
&&\qquad ={1_{\zu\tau_n,\tau_{n+1}\zu}\over\EEE(N-N^{\tau_n})_-}\mal\bigl(V\EEE
(N-N^{\tau_n})\bigr)- \bigl(1_{\zu\tau_n,\tau_{n+1}\zu}V_-\bigr)\mal N
\end{eqnarray*}
is a $\sigma$-martingale, which implies the first claim. The second
follows as in Lemma~~\ref{l:N}.

2. By Lemma~\ref{l:Pstarmartingale} we must show that
$(V+[V,N^\star])Z^{P^\star}$ is a $P$-$\sigma$-martingale. Integration
by parts yields
\[
(V+[V,N^\star])Z^{P^\star}\sim Z^{P^\star}_-\mal \biggl(V+[V,N^\star]+
 \biggl[ V+[V,N^\star],{1\over1+\Delta A^K}\mal M^K \biggr] \biggr).
\]
Hence we must show that the integrator is a $\sigma$-martingale. It equals
\begin{eqnarray*}
&& V+ \biggl[ V,N^\star+{1\over1+\Delta A^K}\mal M^K+
\biggl[N^\star,{1\over1+\Delta A^K}\mal M^K \biggr] \biggr]
\\
&&\qquad =  V+[V,N]+
 \biggl[ V,\tilde a\mal S-K+[\tilde a\mal S,K]-\bigl(\tilde a(1+\Delta
A^K)\bigr)\mal M^{S\star}
\\
&&\hspace*{55.7mm} {}+{1\over1+\Delta A^K}\mal M^K-[\tilde a\mal
M^{S\star},M^K] \biggr].
\end{eqnarray*}
Since $V+[V,N]$ is a $\sigma$-martingale by statement 1, it suffices to
show that
the right-hand side of the long covariation term vanishes.
To this end, observe that using (\ref{e:teil1}), (\ref{e:bdach}) we get
\begin{eqnarray*}
\tilde a\mal S&=&\tilde a\mal M^{S\star}+(\tilde a^\top b^{S\star
})\mal A
\\
&=&\tilde a\mal M^{S\star}+{b^K\over1+\Delta A^K}\mal A
\\
&=&\tilde a\mal M^{S\star}+{1\over1+\Delta A^K}\mal A^K
\end{eqnarray*}
and hence
\begin{eqnarray*}
[\tilde a\mal S,K]
&=&[\tilde a\mal M^{S\star},M^{K}]+[\tilde a\mal M^{S\star},A^{K}]+
 \biggl[{1\over1+\Delta A^K}\mal A^K,K \biggr]
 \\
&=&[\tilde a\mal M^{S\star},M^{K}]+(\tilde a\Delta A^K)\mal M^{S\star}+
{\Delta A^K\over1+\Delta A^K}\mal K.
\end{eqnarray*}
This yields the first claim. The second follows again as in
Lemma~~\ref{l:N}.
\end{pf}

In general we do not know whether $V$ is locally square integrable, or
even special, under $P$. Crucially, this integrability holds under
$P^\star$, which is important for evaluation of the expected squared
hedging error in Section \ref{s:hedge}.

\begin{lemma}\label{l:sub}
We have \textup{1.} $V^2L$, $(v+\vartheta\mal S)^2L$, and
$(v+\vartheta\mal S-V)^2L$ are submartingales for any admissible
endowment/strategy pair $(v,\vartheta)$.

\textup{2.} $V$ is a locally square-integrable semimartingale relative
to $P^\star$.
\end{lemma}

\begin{pf}
1. Let $G:=v+\vartheta\mal S$ and fix $s\leq t$. From Lemmas
\ref{l:L}(3), \ref{l:V} and H\"older's inequality it follows that
\begin{eqnarray*}
(G_s-V_s)^2L_s^2&=& \bigl((G_s-V_s)M^{(s)}_s \bigr)^2
\\
&=&\bigl(E\bigl((G_t-V_t)M^{(s)}_t|\FFF_s\bigr) \bigr)^2
\\
&\leq&E \bigl( (G_t-V_t)^2L_t |\FFF_s \bigr) E \bigl(
\bigl(1-\lambda^{(s)}\mal S_t\bigr)M^{(s)}_t |\FFF_s \bigr)
\\
&=&E \bigl( (G_t-V_t)^2L_t |\FFF_s \bigr)L_s.
\end{eqnarray*}
Integrability follows by setting $t=T$. The claim for $V^2L$ and $G^2L$
follows analogously.

2. For any stopping time $\tau$ we have
\[
E_{P^\star} (V_\tau^2 ) =E (Z^{P\star}_\tau V_\tau^2 ) \leq{E (L_\tau
V_\tau^2 )\over E(L_0)} \leq{E(H^2)\over E(L_0)}
\]
by statement 1, which implies the claim (cf. Lemma~\ref{l:s2}).
\end{pf}

\begin{lemma}\label{l:albert}
Outside some $P\otimes A$-null set we have
\begin{eqnarray}
\label{e:driftrate2} b^{V\star}&=&\tilde c^{VS\star}\tilde a,
\\
\label{e:albert} \tilde c^{S\star}(\tilde c^{S\star})^{-1}\tilde
c^{SV\star}&=&\tilde c^{SV\star}.
\end{eqnarray}
\end{lemma}

\begin{pf}By Lemma~\ref{l:VN} and (\ref{e:teil1}),
(\ref{e:bdach}) we have
\begin{eqnarray*}
0&\sim^\star&V+[V,N^\star]
\\
&\hspace*{-3pt} =&V-\hat a\mal[V,M^{S^\star}]
\\
&\hspace*{-3pt} =&V-\hat a\mal[V,S]+\hat a\mal[V,A^{S^\star}]
\\
&\sim^\star&A^{V\star}-\hat a\mal\langle V,S\rangle^{P^\star} + (\hat
a^\top\Delta A^{S\star} )\mal V
\\
&\sim^\star& (b^{V\star}-\tilde c^{VS\star}\hat a )\mal A +
\bigl((1+\Delta A^K)\tilde a^\top b^{S\star}\Delta A \bigr)\mal A^{V\star }
\\
&\hspace*{-3pt}=& \bigl(b^{V\star}-(1+\Delta A^K)\tilde
c^{VS\star}\tilde a +\Delta A^K b^{V\star} \bigr)\mal A
\\
&\hspace*{-3pt}=& \bigl((1+\Delta A^K)(b^{V\star}-\tilde
c^{VS\star}\tilde a) \bigr)\mal A,
\end{eqnarray*}
which yields the first assertion.

Fix $(\omega,t)\in\Omega\times[0,T]$.
Since
\[
\tilde c_t^{S,V\star}(\omega)= \pmatrix{ \tilde c^{S\star}_t& \tilde
c_t^{SV\star} \cr\noalign{}   (\tilde c_t^{SV\star
})^\top& \tilde c_t^{V\star}} (\omega)
\]
is a symmetric, nonnegative matrix, we have (\ref{e:albert}) by Albert
\cite{albert72}, Theorem~9.1.6.
\end{pf}

The next definition constitutes a first step toward optimal hedging.

\begin{defi}
We call the process
\[
\xi:=(\tilde c^{S\star})^{-1}\tilde c^{SV\star}
\]
\textit{pure hedge coefficient}.
\end{defi}

The following representations of $\xi$
establish the link to (\ref{e:kunitawatanabe}).

\begin{prop}
The pure hedge coefficient $\xi$ satisfies
\begin{equation}\label{e:xi}
\xi\mal\langle S,S\rangle^{P^\star}=
\langle S,V\rangle^{P^\star}
\end{equation}
and
\begin{equation}\label{e:xi2}
\xi\mal\langle M^{S\star},M^{S\star}\rangle^{P^\star}=
\langle M^{S\star},M^{V\star}\rangle^{P^\star}.
\end{equation}
In the univariate case, \textup{(\ref{e:xi})} and
\textup{(\ref{e:xi2})} can be written more plainly as
\begin{equation}\label{e:xi3}
\xi_t={d\langle S,V\rangle^{P^\star}_t
\over d\langle S,S\rangle^{P^\star}_t}
={d\langle M^{S\star},M^{V\star}\rangle^{P^\star}_t
\over d\langle M^{S\star},M^{S\star}\rangle^{P^\star}_t}.
\end{equation}
\end{prop}

\begin{pf}
Lemma~\ref{l:albert} yields
\[
\langle S,V\rangle^{P^\star} =\tilde c^{SV\star}\mal A = (\tilde
c^{S\star}(\tilde c^{S\star})^{-1}\tilde c^{SV\star } )\mal A
=\xi\mal\langle S,S\rangle^{P^\star}.
\]
By (\ref{e:teil3}) and (\ref{e:driftrate2}) we have
\begin{eqnarray}\label{e:MSS}
 \langle M^{S\star},M^{S\star} \rangle^{P^\star}&=&
 \bigl(\tilde c^{S\star}-b^{S\star}(b^{S\star})^\top\Delta A
\bigr)\mal A \nonumber
\\[-8pt]
\\[-8pt]
\nonumber &=& \bigl( (1_d-\tilde c^{S\star}\tilde a\tilde a^\top\Delta
A )\tilde c^{S\star} \bigr)\mal A
\end{eqnarray}
and
\begin{eqnarray}\label{e:MSV}
 \langle M^{S\star},M^{V\star} \rangle^{P^\star}&=&
 \bigl(\tilde c^{SV\star}_t-b^{S\star}_tb^{V\star}_t\Delta A
\bigr)\mal A \nonumber
\\[-8pt]
\\[-8pt]
\nonumber &=& \bigl( (1_d-\tilde c^{S\star}\tilde a\tilde a^\top\Delta
A )\tilde c^{SV\star} \bigr)\mal A,
\end{eqnarray}
where $1_d$ denotes the identity matrix. Equations (\ref{e:MSS}),
(\ref{e:MSV}), (\ref{e:albert}) yield (\ref{e:xi2}).
\end{pf}

The pure hedge coefficient appears in the following decomposition:
\begin{lemma}
There exists a $P^\star$-local martingale $M$
with $M_0=0$ that is \mbox{$P^\star$-}orthogonal
to $M^{S\star}$ (in the sense that $M^{S\star}M$ is a $P^\star$-local
martingale) and such that
\begin{equation}\label{e:fsd}
V=V_0+\xi\mal S+M
\end{equation}
holds.
\end{lemma}

\begin{pf}
By (\ref{e:albert}), (\ref{e:driftrate2}),
(\ref{e:teil3}) we have
\[
0= \bigl(\tilde a^\top\tilde c^{SV\star} -\tilde a^\top\tilde
c^{S\star}(\tilde c^{S\star})^{-1}\tilde c^{SV\star } \bigr)\mal A =
\bigl(b^{V\star}-(b^{S\star})^\top\xi \bigr)\mal A,
\]
which implies that
$M:=V-V_0-\xi\mal S$ is a $P^\star$-$\sigma$-martingale.
By bilinearity and (\ref{e:albert})
the modified second $P^\star$-characteristics of $M$ in the sense of
(\ref{e:mod}) equals
\begin{eqnarray*}
\tilde c^{M\star}&=&\tilde c^{V\star}-2\xi^\top\tilde c^{SV\star }+\xi
^\top \tilde c^{S\star}\xi
\\
&=& \tilde c^{V\star}-2(\tilde
c^{SV\star})^\top(\tilde c^{S\star })^{-1}\tilde c^{SV\star} +(\tilde
c^{SV\star})^\top(\tilde c^{S\star})^{-1}\tilde c^{S\star} (\tilde
c^{S\star})^{-1}\tilde c^{SV\star}
\\
&=&\tilde c^{V\star}-(\tilde
c^{SV\star})^\top(\tilde c^{S\star })^{-1}\tilde c^{SV\star} \leq\tilde
c^{V\star}.
\end{eqnarray*}
Since $V$ is a locally square-integrable semimartingale relative to
$P^\star$, it follows that $M$ is a locally square-integrable
martingale relative to $P^\star$ (cf. \cite{js87}, II.2.29). Since
\[
\langle M^{S\star},M\rangle^{P^\star} =\langle S,V-\xi\mal
S\rangle^{P^\star} = \bigl(\tilde c^{SV\star}-\tilde c^{S\star}(\tilde
c^{S\star })^{-1}\tilde c^{SV\star} \bigr)\mal A=0
\]
by (\ref{e:albert}), we have that $M^{S\star}M$ is a $P^\star$-local
martingale.
\end{pf}

Equation (\ref{e:fsd}) can be interpreted as a process version of the
$P^\star$-F\"ollmer--Schweizer decomposition of $H$. The integrand in
the latter yields the \textit{locally risk-minimizing hedging strategy}
in the sense of \cite{schweizer91} or \cite{foellmerschweizer91}
relative to $P^\star$.

\subsection{\texorpdfstring{Main results.}{Main results}}
\begin{lemma}\label{l:xi}
For any $\FFF_0$-measurable random variable $v$, the feedback equation
\begin{equation}\label{e:phi} \varphi_t=\xi_t- (v+\varphi\mal
S_{t-}-V_{t-}
)\tilde a_t
\end{equation}
has a unique solution $\varphi(v):=\varphi\in L(S)$.
\end{lemma}

\begin{pf}
In the proof of Theorem~\ref{t:hedge} below it is shown that $\xi\in
L(S)$. The stochastic differential equation
\begin{equation}\label{e:G}
G= \bigl(\xi-(v-V_{-})\tilde a \bigr)\mal S - G_-\mal(\tilde a\mal S)
\end{equation}
has a unique solution $G$ by Jacod \cite{jacod79}, (6.8). If we set
$\varphi_t:=\xi_t- (v+G_{t-}-V_{t-} )\tilde a_t$, then $\varphi\in
L(S)$ solves (\ref{e:phi}).

If, on the other hand, some $\tilde\varphi\in L(S)$ solves
(\ref{e:phi}) as well, then $\widetilde G:=\tilde\varphi\mal S$ is a
solution to (\ref{e:G}). This implies $\widetilde G=G$ and hence
$\tilde\varphi=\varphi$.
\end{pf}

We are now ready to state our first main result.

\begin{teo}\label{t:hedge}
\textup{1.} The process $\varphi:=\varphi(v_0)$ given by the feedback
expression \textup{(\ref{e:phi})} is an optimal hedging strategy for
initial endowment $v_0\in L^2(\Omega ,\FFF_0,P)$.

\textup{2.} $(v_0,\varphi):=(V_0,\varphi(V_0))$ is an optimal
endowment/strategy pair.
\end{teo}

\begin{pf}
1. Denote by
\[
\label{e:bc} \left(\pmatrix{b^S\cr   b^V\cr   b^K} , \pmatrix{ c^{S} &
c^{SV} & c^{SK}\cr  c^{VS} & c^V & c^{VK} \cr  c^{KS} & c^{KV} & c^{K}}
, F^{S,V,K},A \right)
\]
$P$-differential characteristics of $(S,V,K)$ relative to the
``truncation'' function $h(x,z,y):=(x,z1_{\{|z|\leq1\}},y)$ on
$\rr^d\times\rr\times\rr$. [Should $V$ be a $P$-special semimartingale,
we could also choose the identity $h(x,z,y)=(x,z,y)$ as usual in this
paper.] Along the same lines as in the proof of
Lemma~\ref{l:SunterPstern} it follows that
\begin{eqnarray}
\label{e:cSV} \tilde c^{SV\star}&=& {1\over1+\Delta A^K}
\biggl(c^{SV}+\int xz(1+y)F^{S,V,K}\bigl(d(x,z,y)\bigr) \biggr),
\\
\label{e:cV} \tilde c^{V\star} &=&{1\over1+\Delta A^K}
\biggl(c^{V}+\int z^2(1+y)F^{S,V,K}\bigl(d(x,z,y)\bigr) \biggr).
\end{eqnarray}

Let $\bar\varphi$ be an optimal hedging strategy for initial endowment
$v_0$, denote by $G:=v_0+\bar\varphi\mal S$ its value process, and set
$\bar\xi:=\bar\varphi+(G_--V_-)\tilde a$. Moreover, let
$\vartheta\in\overline\Theta$ and $\widetilde G:=\vartheta \mal S$. In
the proof of Lemma~\ref{l:V} it is shown that $Z\widetilde G$ is a
martingale for $Z:=(G-V)L$. Integration by parts yields $Z\widetilde
G=L_0\EEE(K)(G-V)\widetilde G=L_-\mal U$ with
\begin{eqnarray*}
U&=&(G-V)\widetilde G+\bigl((G-V)\widetilde G\bigr)_-\mal
K+[(G-V)\widetilde G,K]
\\
&=&(G-V)_-\mal\widetilde G+\widetilde
G_-\mal(G-V)+[G-V,\widetilde G] + \bigl((G-V)\widetilde G\bigr)_-\mal K
\\
&&{}+(G-V)_-\mal[\widetilde G,K]+\widetilde G_-\mal
[G-V,K]+\bigl[G-V,[\widetilde G,K]\bigr]
\\
&=&(G-V)_-\mal \bigl(\widetilde G_-\mal N+\vartheta\mal(S+[S,N]) \bigr)
\\
&&{}+\bar\xi\mal \bigl(\widetilde G_-\mal(S+[S,K])+ \bigl[\widetilde
G,S+[S,K] \bigr] \bigr)
\\
&&{}-\widetilde G_-\mal(V+[V,K])- \bigl[\widetilde G,V+[V,K] \bigr].
\end{eqnarray*}
The first term on the right-hand side is a $\sigma$-martingale by
Lemma~\ref{l:N}. By (\ref{e:N}), the remaining two terms equal
\begin{eqnarray*}
&& G_-\mal \bigl(\bar\xi\mal(S+[S,N])-(V+[V,N]) \bigr)
\\
&&\qquad {} +(\tilde aG_-+\vartheta)\mal \bigl(\bar\xi\mal
\bigl[S,S+[S,K] \bigr]- \bigl[V,S+[S,K] \bigr] \bigr).
\end{eqnarray*}
The first line is a $\sigma$-martingale by Lemmas \ref{l:N} and
\ref{l:VN}. By (\ref{e:querc2}), (\ref{e:cdach}) we have
\begin{eqnarray}\label{e:sssk}
\nonumber \bigl[S,S+[S,K]\bigr]&=&[S^{c},S^{c}]+\int xx^\top(1+y)\mu
^{S,K}\bigl(d(x,y)\bigr)
\\
&\sim& \biggl(c^S+\int xx^\top(1+y)F^{S,K}\bigl(d(x,y)\bigr)
\biggr)\mal A \nonumber
\\[-8pt]
\\[-8pt]
\nonumber &=&\bar c\mal A
\\
\nonumber &=&  \bigl((1+\Delta A^K)\tilde c^{S\star} \bigr)\mal A.
\end{eqnarray}
Similarly, (\ref{e:cSV}) yields
\begin{eqnarray}\label{e:vssk}
\bigl[S+[S,K],V\bigr]&\sim&
 \biggl(c^{SV}+\int xz(1+y)F^{S,V,K}\bigl(d(x,z,y)\bigr) \biggr)\mal A
 \nonumber
 \\[-8pt]
 \\[-8pt]
 \nonumber
 &=& \bigl((1+\Delta A^K)\tilde c^{SV\star} \bigr)\mal A.
\end{eqnarray}
For later use, we observe that
\begin{equation}\label{e:vvkv}
\bigl[V+[V,K],V\bigr]\sim\bigl((1+\Delta A^K)\tilde
c^{V\star}\bigr)\mal A 1
\end{equation}
by (\ref{e:cV}). Altogether, we have that $ ((\tilde
aG_-+\vartheta)(1+\Delta A^K)(\tilde c^{S\star}\bar\xi -\tilde
c^{SV\star}) )\mal A$ is a $\sigma$-martingale. This being true for any
$\vartheta$, we have
\begin{equation}\label{e:xiquer}
\tilde c^{S\star}\bar\xi-\tilde c^{SV\star}=0
\end{equation}
$P\otimes A$-almost everywhere.\vadjust{\goodbreak}

For $n\in\nn$ define the predictable set $D_n:=\{|\xi|\vee|\bar\xi
|\leq n\}$. Corollary \ref{co:adjustment2} and (\ref{e:xiquer}),
(\ref{e:albert}) yield
\[
 \bigl((\bar\xi-\xi)1_{D_n} \bigr)\mal A^{S\star}
= \bigl(\bigl(\bar\xi-(\tilde c^{S\star})^{-1}\tilde
c^{SV\star}\bigr)^\top \tilde c^{S\star}\tilde a1_{D_n} \bigr)\mal A =0
\]
as well as
\begin{eqnarray*}
&& \bigl\langle\bigl((\bar\xi-\xi)1_{D_n}\bigr)\mal M^{S\star},
\bigl((\bar\xi-\xi)1_{D_n}\bigr)\mal M^{S\star} \bigr\rangle^{P\star}
\\
&&\qquad = \bigl((\bar\xi-\xi)^\top\hat c^{S\star}(\bar\xi-\xi )1_{D_n} \bigr)\mal A
\\
&&\qquad \leq \bigl((\bar\xi-\xi)^\top\tilde c^{S\star}(\bar\xi-\xi
)1_{D_n} \bigr)\mal A=0.
\end{eqnarray*}
Consequently, $((\bar\xi-\xi)1_{D_n})\mal S=0$ for any $n$, which in
turn implies $\bar\xi-\xi\in L(S)$ and $(\bar\xi-\xi )\mal S=0$ by
Lemma~\ref{l:easily}. In particular, we have
$\xi=\bar\xi-(\bar\xi-\xi)\in L(S)$. The proof of Lemma~\ref{l:xi}
yields that $\varphi\mal S=\bar\varphi\mal S$ as well. In particular,
$\varphi$ is admissible and an optimal hedging strategy for initial
endowment $v_0$.

2. This follows essentially as statement 1. We only have to determine
the optimal initial endowment. Denote by $(v_0,\bar\varphi)$ an optimal
endowment/strategy pair and let $Z$ be as in the first part of the
proof. Parallel to (\ref{e:optimality}), we conclude that $E(vZ_T)=0$
for any $v\in L^2(\Omega,\FFF_0,P)$, which implies
$0=E(Z_T|\FFF_0)=Z_0=L_0(v_0-V_0)$. Consequently, $v_0=V_0$ as claimed.
\end{pf}

As is well known, the gains process $\varphi\mal S$ can be expressed
more explicitly.

\begin{cor}
The gains process of the optimal hedge in
Theorem~\textup{\ref{t:hedge}} equals
\[
\varphi\mal S=\EEE(-\tilde a\mal S) \biggl({\xi+(V_--v_0)\tilde
a\over\EEE(-\tilde a\mal S)_-}\mal
 \biggl(S+{\tilde a\over1-\tilde a^\top\Delta S}\mal[S,S]
\biggr) \biggr)
\]
unless $\EEE(-\tilde a\mal S)$ jumps to \textup{0}.
\end{cor}

\begin{pf}
By \cite{jacod79}, (6.8) the stochastic differential equation
$X=Y+X_-\mal Z$ with two semimartingales $Y,Z$ such that $Y_0=0$ is
uniquely solved by
\[
X=\EEE(Z) \biggl({1\over\EEE(Z)_-}\mal Y-{1\over\EEE(Z)}\mal[Y,Z]
\biggr)
\]
unless $\EEE(Z)$ jumps to 0. Since
\[
\varphi\mal S=\bigl(\xi-(v_0-V_-)\tilde a\bigr)\mal S+(\varphi\mal
S)_-\mal(-\tilde a\mal S),
\]
the assertion follows.
\end{pf}

Finally, we state formulas for the hedging error.

\begin{teo}\label{t:error}
The expected squared hedging error of the optimal hedge in
Theorem\textup{ \ref{t:hedge}} equals
\begin{eqnarray}
\nonumber && E \bigl((v_0+\varphi\mal S_T-H)^2 \bigr)
\\
\nonumber &&\qquad = E \bigl((v_0-V_0)^2L_0+ \bigl( \bigl(\tilde c^{V\star}
-(\tilde c^{SV\star})^\top(\tilde c^{S\star})^{-1}\tilde c^{SV\star }
\bigr)L \bigr)\mal A_T \bigr)
 \\
\nonumber &&\qquad=E \bigl((v_0-V_0)^2L_0+L\mal (\langle V,V\rangle^{P^\star}
-\xi\mal\langle V,S\rangle^{P^\star} )_T \bigr)
\\
\label{e:pfehler}
&&\qquad= E \bigl((v_0-V_0)^2L_0+L\mal\langle V-\xi\mal
S,V-\xi\mal S\rangle ^{P^\star}_T \bigr)
\\
\nonumber &&\qquad= E_{P^\star} \bigl((v_0-V_0)^2
\\
\nonumber &&\phantom{\qquad= E_{P^\star} \bigl(}
{} + \bigl( \bigl(\tilde
c^{V\star} -(\tilde c^{SV\star})^\top(\tilde
c^{S\star})^{-1}\tilde c^{SV\star } \bigr)\EEE (A^K) \bigr)\mal A_T \bigr)E(L_0)
\\
\label{e:psternfehler} &&\qquad=E_{P^\star}
\bigl((v_0-V_0)^2+\EEE(A^K)\mal\langle V-\xi\mal S,V-\xi \mal
S\rangle^{P^\star}_T \bigr)E(L_0).
\end{eqnarray}
\end{teo}

\begin{pf}In view of Proposition~\ref{p:spitzklammer}, the
second equality is obvious. The third and the last follow from
\[
\label{e:VxiS} \langle S,V-\xi\mal S\rangle^{P^\star} = \bigl(\tilde
c^{SV\star}-\tilde c^{S\star}(\tilde c^{S\star })^{-1}\tilde
c^{SV\star} \bigr)\mal A=0.
\]

Define $G:=v_0+\varphi\mal S$ and $Z:=(G-V)L$ as in the proof of
Theorem~\ref{t:hedge}. Since $(G-V)^2L$ is a submartingale by
Lemma~\ref{l:sub}, there exists a unique increasing predictable process
$B$ with $B_0=0$ and such that $(G-V)^2L-B$ is a martingale. Since
\[
E \bigl((v_0+\varphi\mal S_T-H)^2 \bigr)= E\bigl((G_T-V_T)^2L_T\bigr)
=E\bigl((G_0-V_0)^2L_0\bigr)+E(B_T),
\]
the first equality in the theorem holds if
\begin{equation}\label{e:B}
B= \bigl((\tilde c^{V\star}-\xi^\top\tilde c^{SV\star})L \bigr)\mal A.
\end{equation}
Since $GZ$ and $Z$ are martingales, we have
\begin{eqnarray}\label{e:VZ}
\nonumber && -(G-V)^2L\sim VZ
\\
&&\qquad \sim Z_-\mal V+[V,Z]
\nonumber
\\[-8pt]
\\[-8pt]
\nonumber
&&\qquad = \bigl((G-V)_-L_- \bigr)\mal V
\\
\nonumber &&\qquad\quad{} + \bigl[V,(G-V)_-\mal L+L_-\mal(G-V)+[G-V,L] \bigr].
\end{eqnarray}
In view of
\[
G-V=v_0+\xi\mal S-\bigl((G-V)_-\tilde a\bigr)\mal S-V
\]
(\ref{e:VZ}) equals
\begin{eqnarray*}
&& \bigl((G-V)_-L_- \bigr)\mal \bigl(V+ \bigl[V,K-\tilde a\mal
S-[\tilde a\mal S,K] \bigr] \bigr)
\\
&&\qquad{} +L_-\mal \bigl[V+[V,K],\xi\mal S-V
\bigr].
\end{eqnarray*}
By Lemma~\ref{l:VN} the first term is a $\sigma$-martingale and hence
\begin{eqnarray*}
(G-V)^2L&\sim& -L_-\mal \bigl(\xi\mal \bigl[V+[V,K],S
\bigr]-\big[V+[V,K],V\big ] \bigr)
\\
&=&-L_-\mal \bigl(\xi\mal\big[V,S+[S,K]\big]-\big[V+[V,K],V\big ]
\bigr)
\\
&\sim&  \bigl(L_-(1+\Delta A^K) (\tilde c^{V\star}-\xi^\top\tilde
c^{SV\star}) \bigr)\mal A
\\
&=& \bigl((L-L_-\Delta M^K)(\tilde c^{V\star}-\xi^\top\tilde c^{SV\star
}) \bigr)\mal A
\end{eqnarray*}
by (\ref{e:vssk}) and (\ref{e:vvkv}). Since $\Delta M^K\mal
U=[M^K,U]=\Delta U\mal M^K$ is a $\sigma$-martingale for any
predictable process $U$ of finite variation (cf. \cite{js87}, I.4.49),
we obtain
\[
(G-V)^2L \sim \bigl(L(\tilde c^{V\star}-\xi^\top\tilde c^{SV\star})
\bigr)\mal A.
\]
Therefore the difference of both sides of (\ref{e:B}) is a predictable
$\sigma$-martingale of finite
variation and hence 0.

It remains to be shown that (\ref{e:psternfehler}) equals (\ref{e:pfehler}).
Integration by parts yields
\begin{eqnarray*}
&& Z^{P^\star} \bigl(E(L_0)\EEE(A^K)\mal\langle V-\xi\mal S,V-\xi \mal
S\rangle^{P^\star} \bigr)
\\
&&\qquad = \bigl(Z^{P^\star}E(L_0)\EEE(A^K) \bigr)\mal\langle V-\xi\mal
S,V-\xi \mal S\rangle^{P^\star}
\\
&&\qquad\quad{}+ \bigl(E(L_0)\EEE(A^K)\mal\langle
V-\xi\mal S,V-\xi\mal S\rangle ^{P^\star} \bigr)_-\mal Z^{P^\star}
\\
&&\qquad =L\mal\langle V-\xi\mal S,V-\xi\mal S\rangle^{P^\star}+M
\end{eqnarray*}
with some $P$-local martingale $M$.
Hence
\begin{eqnarray*}
&& E_{P^\star} \bigl(\EEE(A^K)\mal \langle V-\xi\mal S,V-\xi\mal
S\rangle^{P^\star}_{T_n} \bigr)E(L_0)
\\
&&\qquad = E \bigl(Z^{P^\star}_{T_n}
 \bigl(E(L_0)\EEE(A^K)\mal\langle V-\xi\mal S,V-\xi\mal S\rangle
^{P^\star }_{T_n} \bigr) \bigr)
\\
&&\qquad = E (L\mal\langle V-\xi\mal S,V-\xi\mal S\rangle^{P^\star
}_{T_n} ),
\end{eqnarray*}
where $(T_n)_{n\in\nn}$ denotes a localizing sequence for $M$. Monotone
convergence yields that (\ref{e:psternfehler}) equals
(\ref{e:pfehler}).
\end{pf}

If the results in this paper are to be applied to concrete models, it
is not necessary to determine all the processes that have been
introduced. Instead, one may proceed as follows: first one determines
the opportunity process $L$ and the adjustment process $\tilde a$ using
the characterization in Theorem~\ref{t:unique}. These processes yield
the modified mean-variance tradeoff process $K$, the
opportunity-neutral measure $P^\star$ and the variance-optimal
logarithm process $N$. Finally, the mean-value process $V$ leads to the
pure hedge coefficient $\xi$ and hence to the optimal hedge $\varphi$.

\subsection{\texorpdfstring{Connections to the literature.}{Connections to the literature}}\label{su:link}
In this section we clarify the link of our results to the literature.
If $S$ is a martingale, we are in the setup of F\"ollmer and Sondermann
\cite{foellmersondermann86}. In our notation, they show that the
optimal hedge $\varphi$ satisfies
\begin{equation}\label{e:fs}
\varphi_t={d\langle S,V\rangle_t\over d\langle S,S\rangle_t},
\end{equation}
where
\begin{equation}\label{e:Vfs}
V_t=E(H|\FFF_t).
\end{equation}
Applying our results to the martingale case, one immediately verifies
that $L=1$, $\tilde a=0$, $K=0$, $N=0$, $Q^\star=P^\star=P$.
Consequently, equation (\ref{e:V}) for the mean-value process of the
option reduces to (\ref{e:Vfs}). Moreover, the optimal hedge $\varphi$
coincides with the pure hedge $\xi$, which satisfies $\xi\mal\langle
S,S\rangle=\langle S,V\rangle$ in accordance with~(\ref{e:fs}).

Schweizer \cite{schweizer94} goes beyond the martingale case. He shows
that if the MVT process $\widehat{K}$ is deterministic, then the
optimal hedging strategy for initial endowment $v_0$ contains a
feedback element and is of the form
\begin{equation}\label{e:s94hedge}
\varphi_t =\xi_t-(v_0+\varphi\mal S_{t-}-V_{t-})\tilde{\lambda}_t
\end{equation}
with $\tilde\lambda$ from (\ref{e:modlambda}). Here, the \textit{pure
hedge coefficient} $\xi$ is the integrand in the
\textit{F\"ollmer--Schweizer decomposition} of the claim, that is,
\[
H=V_0+\xi\mal S_T+R_T,
\]
where $V_0$ is a $\FFF_0$-measurable random variable and $R$ denotes a
martingale which is orthogonal to $M^S$ (in the sense that $M^{S}R$ is
a local martingale). In order to express the pure hedge coefficient
similarly as in (\ref{e:fs}), recall that the minimal signed martingale
measure $Q$ is given by
\[
{dQ\over dP}:=\EEE(-\hat\lambda\mal M^S)_T.
\]
If we define $V$ as ``$Q$-conditional expectation'' of $H$ in the sense of
\begin{equation}\label{e:s94price}
V_t:=E \bigl( H\EEE\bigl(\bigl(-\hat\lambda1_{\zu t,T\zu}\bigr)\mal
M^S\bigr)_T |\FFF_t \bigr),
\end{equation}
then the pure hedge coefficient can be written as
\begin{equation}\label{e:s94pure}
\xi_t={d\langle S,V\rangle_t\over d\langle S,S\rangle_t}
={d\langle M^S,M^V\rangle_t\over d\langle M^S,M^S\rangle_t}.
\end{equation}
The hedging error satisfies the equation
\begin{eqnarray}\label{e:s94error}
&& E \bigl((v_0+\varphi\mal S_T-H)^2 \bigr) \nonumber
\\[-8pt]
\\[-8pt]
\nonumber &&\qquad =E \bigl((v_0-V_0)^2 +\EEE(\widehat K)\mal\langle
V-\xi\mal S,V-\xi\mal S\rangle_T \bigr){1\over\EEE (\widehat K)_T} .
\end{eqnarray}
In these formulas, all predictable covariation processes refer to the
original probability measure $P$.

It is easy to see that (\ref{e:s94hedge})--(\ref{e:s94error}) are special
cases of our general results. To this end, recall
that $L={\EEE(\widehat K)/\EEE(\widehat K)_T}$, $P^\star=P$, and $\tilde a=\tilde \lambda$ in the
case of deterministic MVT (cf. Corollaries \ref{co:DMVT} and
\ref{co:MMM}). Hence
\[
N^\star=-\bigl((1+\Delta A^K)\tilde a\bigr)\mal M^{S\star}
=-\bigl((1+\Delta\widehat K)\tilde\lambda\bigr)\mal
M^{S}=-\hat\lambda\mal M^S.
\]
Consequently, (\ref{e:s94hedge}), (\ref{e:s94price}),
(\ref{e:s94pure}), (\ref{e:s94error}) correspond to (\ref{e:phi}),
(\ref{e:V2}), (\ref{e:xi3}), (\ref{e:psternfehler}), respectively.

If the MVT process fails to be deterministic, the above formulas do not
lead to the optimal hedge any more. Following Hipp \cite{hipp93},
\cite{schweizer96} observes that a key role in the general case is
played by the variance optimal signed martingale measure $Q^\star$ and
the adjustment process $\tilde a$. Schweizer characterizes both the
adjustment process and the optimal hedging strategy in terms of
backward stochastic differential equations. The use of these BSDEs in
practice is complicated by their involved boundary conditions, which
themselves depend on the unknown solution.

Rheinl\"ander and Schweizer \cite{rheinlaenderschweizer97} show that
the optimal hedging strategy $\varphi$ satisfies similar equations as
in the case of deterministic MVT \textit{if $S$ is continuous}. More
specifically, it is of feedback form
\[
\label{e:feedbackform} \varphi_t =\xi_t-(v_0+\varphi\mal
S_{t-}-V_{t-})\tilde a_t,
\]
where $V_t:=E_{Q^\star}(H|\FFF_t)$ is the martingale generated by $H$
relative to the variance-optimal S$\sigma$MM 
$Q^\star$ and the \textit{pure hedge coefficient} $\xi$ is the
integrand in the Galtchouk--Kunita--Watan\-abe decomposition of $H$
relative to $Q^\star$ rather than $P$, that is,
\[
\xi_t={d\langle S,V\rangle^{Q^\star}_t\over d\langle S,S\rangle
^{Q^\star}_t}.
\]
This equation corresponds to our expression (\ref{e:xi3}) because
the predictable covariation does not depend on the probability measure
for continuous processes.

An alternative approach in the continuous case is pursued by
Gourieroux, \mbox{Laurent} and Pham \cite{gourierouxal98} who use a new
numeraire $\EEE ( -\tilde {a}\mal S )$ combined with a change of
measure to transform the original semimartingale problem to a
martingale problem \`{a} la F\"ollmer and Sondermann
\cite{foellmersondermann86}. The task of computing $\tilde a$ has
become a separate issue in the literature. It is tackled in a number of
diffusion or jump-diffusion settings, for example, by Laurent and Pham
\cite{laurentpham99}, Biagini, Guasoni and Pratelli~\cite{biaginial00},
Biagini and Guasoni \cite{biaginiguasoni02}, Hobson \cite{hobson04}.
Our characterization of the adjustment process in
Theorem~\ref{t:unique} appears to be more suitable for direct
computations than the methods available to date (cf.
\cite{cernykallsen06awp}).

The literature on discontinuous processes is more limited. Two partial
results are reported by Arai \cite{arai05} and Lim \cite{lim05}. Arai
extends the numeraire method of Gourieroux, Laurent and Pham
\cite{gourierouxal98} to discontinuous semimartingales assuming that
$Q^{\star}$ is equivalent to $P$ and shows that $V$ in
(\ref{e:feedbackform}) is a $Q^{\star}$-martingale. However, Arai's
results are hard to use for explicit computations since he does not
provide a method for obtaining $\tilde{a}$.

Lim \cite{lim05} uses BSDEs to compute the optimal hedge in a jump
diffusion setting where asset price characteristics are adapted to a
Brownian filtration. In addition he requires a certain
\textit{martingale invariance property}. He characterizes the optimal
hedge explicitly at the cost of a somewhat restrictive model setup.

Finally, we want to explain another close link of our results to the
formulas (\ref{e:s94hedge}--\ref{e:s94error}) of \cite{schweizer94}.
We already observed in Lemma~\ref{l:VOMMMMM} that the variance-optimal
S$\sigma$MM 
$Q^\star$ is the minimal S$\sigma$MM 
relative to $P^\star$. Moreover, $\tilde a$ and $\hat a$ coincide with
the processes $\tilde\lambda$ and $\hat\lambda$ in \cite{schweizer94}
or Section \ref{su:ppstar} relative to $P^\star$ instead of $P$.
Consequently, equations (\ref{e:phi}), (\ref{e:V2}), (\ref{e:xi3}) are
$P^\star$-versions of the formulas (\ref{e:s94hedge}),
(\ref{e:s94price}), (\ref{e:s94pure}). The change of measure $P\to
P^\star$ neutralizes the effect of stochastic mean-variance tradeoff
which makes the results in \cite{schweizer94} break down. With the
hedging error one has to be slightly more careful. Since $\widehat
K^\star=A^K$, we can view (\ref{e:psternfehler}) essentially as a
$P^\star$-version of (\ref{e:s94error}). We only have to replace the
deterministic second factor $1/\EEE(\widehat K)_T$ by
\[
E \biggl({1\over\EEE(\widehat K^\star)_T} \biggr) =E
\biggl({1\over\EEE(A^K)_T} \biggr) =E_{P^\star} \biggl({E(L_0)\over
L_T} \biggr)=E(L_0).
\]

\begin{appendix}
\section*{\texorpdfstring{APPENDIX}{Appendix}}\label{app}

\renewcommand{\thesubsection}{\Alph{section}.\arabic{subsection}}
\renewcommand{\theequation}{\Alph{section}.\arabic{equation}}
\setcounter{teo}{0}

\subsection{\texorpdfstring{Locally square-integrable
semimartingales.}{Locally square-integrable semimartingales}}

\begin{defi}\label{d:s2}
For any special semimartingale $X$ we define
\[
\|X\|_{\SSS}:=\|X_0\|_2+\|\sqrt{[M^X,M^X]_T}\|_2+\|\mathrm
{var}(A^X)_T\|_2,
\]
where $\operatorname{var}(A^X)$ denotes the variation process of $A^X$
and $\|\cdot\| _2$ the $L^2$-norm. $X$ is said to belong to the set
$\SSS$ of \textit{square-integrable semimartingales} if
$\|X\|_{\SSS}<\infty$. The elements of the corresponding localized
class $\ssl$ are called \textit{locally square-integrable
semimartingales}.
\end{defi}

\begin{lemma}\label{l:s2}
For any semimartingale $X$, we have equivalence between:
\begin{longlist}[4.]
\item[1.] $X\in\ssl$.

\item[2.] $X_0\in L^2(P)$ and $X$ is a locally square-integrable
semimartingale in the sense of \textup{\cite{js87}, II.2.27,} that is,
it is a special semimartingale whose local martingale part is locally
square-integrable.

\item[3.] $X$ is \textit{locally in $L^2$} in the sense of
\textup{\cite{delbaenschachermayer96}}, that is, it belongs locally to
the class of processes $Y$ with
\[
\sup \{E(Y_\tau^2)\dvtx \tau\mbox{ finite stopping time} \} <\infty.
\]

\item[4.] $X$ belongs locally to the class of processes $Y$ satisfying
$E(Y_\tau^2)<\infty$ for any finite stopping time $\tau$.
\end{longlist}
\end{lemma}

\begin{pf}
We refer to the time set $\rp$ rather than
$[0,T]$ in this proof.
\begin{longlist}[1${}\Rightarrow{}$2:]
\item[1${}\Rightarrow{}$2:] This follows from \cite{js87}, II.2.28 and
from the inequality
\[
E \biggl(\,\sup_{t\in\rp}(Y_t-Y_0)^2 \biggr)\leq8\|Y\|_{\SSS}^2,
\]
which holds for any semimartingale $Y$ (cf. \cite{protter04},
Theorem~IV.5).

\item[2${}\Rightarrow{}$3:] This follows immediately from \cite{js87},
II.2.28.

\item[3${}\Rightarrow{}$4:] This is trivial.

\item[4${}\Rightarrow{}$1:] Define a sequence of stopping times
$\tau_n:=\inf\{t\in\rp\dvtx |X_t|>n\} \wedge n$. Since
$\sup_{t\in\rp}|X^{\tau_n}_t|\leq n+|X_{\tau_n}|$ is integrable, $X$ is
a special semimartingale (cf. \cite{js87}, I.4.23). Choose a localizing
sequence $(\sigma_n)_{n\in\nn}$ for the locally bounded process
$\operatorname{var}(A^X)$. Then
\begin{eqnarray*}
\sup_{t\in\rp} |(M^X)^{\sigma_n\wedge\tau_n}_t |^2 &\leq&
3\sup_{t\in\rp} |X^{\tau_n}_t |^2+3\sup_{t\in \rp} |(A^X)^{\sigma_n}_t
|^2 +3|X_0|^2
\\
&\leq&6n^2+6 |X_{\tau_n} |^2+3\sup_{t\in\rp} (\operatorname{var}
(A^X)^{\sigma_n}_t )^2+3|X_0|^2
\end{eqnarray*}
is integrable for any $n\in\nn$, which yields $X\in\ssl$ (cf.
\cite{js87}, I.4.50c).\quad\qed
\end{longlist}\noqed
\end{pf}

The following result on square integrability of exponential
semimartingales is needed in the proof of Proposition~\ref{p:PP*}. It
extends a parallel statement for local martingales in \cite{jacod79},
(8.27).

\begin{lemma}\label{l:jacod}
Let $X$ be a locally square-integrable semimartingale such that
$\langle M^X,M^X\rangle$ and the variation process $\operatorname{var}
(A^X)$ are bounded. Then
\[
E \biggl(\sup_{t}\EEE(X)^2_t \biggr)<\infty.
\]
\end{lemma}

\begin{pf}
For ease of notation we prove the assertion for the time set $\rp$
rather than $[0,T]$. Denote by $m\in\rp$ an upper bound of $V:=\langle
M^X,M^X\rangle +\operatorname{var}(A^X)$. We write $Z:=\EEE(X)$ and
$Y^*_t:=\sup_{s\in[0,t]}|Y_s|$ for any process $Y$. For $n\in\nn$
define stopping times
\[
\sigma_n:=\inf\{t\in\rp\dvtx |Z_t|\geq n\}.
\]
Fix $n$ and set $\widetilde Z:=Z^{\sigma_n}$.

\textit{Step} 1: We show that
\[
E ((\widetilde Z^*_{\tau_-})^2 )\leq 3+(12+3m)E \bigl((\widetilde
Z_-^2\wedge n^2)\mal V_{\tau-} \bigr)
\]
for any predictable stopping time $\tau$.

In view of $\widetilde Z=1+(\widetilde Z_-1_{\auf0,\sigma_n\zu})\mal
M^X+(\widetilde Z_-1_{\auf 0,\sigma_n\zu})\mal A^X$, we have
\[
E ((\widetilde Z^*_{\tau_-})^2 ) \leq3+3E \bigl(
\bigl(\bigl(\bigl(\widetilde Z_-1_{\auf0,\sigma_n\zu}\bigr)\mal
M^X\bigr)^*_{\tau _-} \bigr)^2 \bigr) +3E \bigl(
\bigl(\bigl(\bigl(\widetilde Z_-1_{\auf0,\sigma_n\zu}\bigr)\mal
A^X\bigr)^*_{\tau _-} \bigr)^2 \bigr).
\]
Since $\tau$ is predictable, Doob's inequality yields
\begin{eqnarray*}
E \bigl( \bigl(\bigl(\bigl(\widetilde
Z_-1_{\auf0,\sigma_n\zu}\bigr)\mal M^X\bigr)^*_{\tau _-} \bigr)^2
\bigr) &\leq& 4E \bigl(\bigl(\widetilde
Z_-^21_{\auf0,\sigma_n\zu}\bigr)\mal \langle M^X,M^X\rangle_{\tau_-} \bigr)
\\
&\leq&4E \bigl((\widetilde Z_-^2\wedge n^2)\mal V_{\tau_-} \bigr).
\end{eqnarray*}
For the part of finite variation we have
\begin{eqnarray*}
 \bigl(\bigl(\bigl(\widetilde Z_-1_{\auf0,\sigma_n\zu}\bigr)\mal A^X\bigr)^*_{\tau
_-} \bigr)^2 &\leq&  \bigl((|\widetilde Z_-|\wedge
n)\mal\operatorname{var}(A^X)_{\tau _-} \bigr)^2
\\
&\leq& (\widetilde Z^2_-\wedge n^2)\mal\operatorname{var}(A^X)_{\tau_-}
\mathrm {var}(A^X)_\infty
\end{eqnarray*}
and hence
\[
E \bigl( \bigl(\bigl(\bigl(\widetilde
Z_-1_{\auf0,\sigma_n\zu}\bigr)\mal A^X\bigr)^*_{\tau _-} \bigr)^2
\bigr) \leq mE \bigl((\widetilde Z_-^2\wedge n^2)\mal V_{\tau_-}
\bigr).
\]

\textit{Step} 2: For $\vartheta\in\rp$ define the predictable stopping
time $T_\vartheta:=\inf\{t\in\rp\dvtx V_t\geq\vartheta\}$ (cf.
\cite{js87}, I.2.13). Step 1 yields that
\[
f(\vartheta):=E \bigl((\widetilde Z^*_{T_\vartheta-}\wedge n)^2 \bigr)
\leq3+(12+3m)E \bigl((\widetilde Z_-^2\wedge n^2)\mal V_{T_\vartheta -}
\bigr).
\]
Since $\vartheta\mapsto T_\vartheta$ is the pathwise generalized
inverse of $V$, we have
\[
(\widetilde Z_-^2\wedge n^2)\mal V_{T_\vartheta-}
=\int_0^{V_{T_\vartheta-}}(\widetilde Z^2_{T_\varrho-}\wedge
n^2)\,d\varrho \leq\int_0^\vartheta(\widetilde Z^*_{T_\varrho-}\wedge
n)^2\,d\varrho
\]
and hence
\[
f(\vartheta)\leq3+(12+3m)\int_0^\vartheta f(\varrho)\,d\varrho
\]
for any $\vartheta\in\rp$. By Gronwall's inequality this implies
$f(\vartheta)\leq3e^{(12+3m)\vartheta}$. Since $T_{m+1}=\infty$, we
have
\[
E \biggl(n^2\wedge\sup_{t\leq\sigma_n}Z^2_t \biggr)=E \bigl((\widetilde
Z^*_{\infty -}\wedge n)^2 \bigr) \leq3e^{(12+3m)(m+1)}.
\]
The assertion follows now from monotone convergence.
\end{pf}

\subsection{\texorpdfstring{$\sigma$-martingales.}{$\sigma$-martingales}}\label{su:sigma}
The following facts on $\sigma$-martingales and integrability can be
found, for example, in \cite{kallsen03}. We summarize them here for the
convenience of the reader.

\begin{defi}\label{d:sigmamartingal}
A semimartingale $X$ is called \textit{$\sigma$-martingale} if there
exists an increasing sequence $(D_n)_{n\in\nn}$ of predictable sets
such that $D_n\uparrow\Omega\times\rp$ up to an evanescent set and
$1_{D_n}\mal X$ is a uniformly integrable martingale for any $n\in
\nn$.
\end{defi}

\begin{rem}\label{r:sigmamartingal}
\textit{Uniformly integrable martingale} can be replaced by
\textit{local martingale} in the previous definition.
\end{rem}

\begin{lemma}\label{l:sigmamartingale}
Let $X$ be a semimartingale with differential characteristics
$(b,c,F,A)$ relative to some truncation function $h$. Then $X$ is a
\mbox{$\sigma$-}martingale if and only if
$\int_{\{|x|>1\}}|x|F(dx)<\infty$ and
\[
b+\int\bigl(x-h(x)\bigr)F(dx)=0
\]
hold outside some $P\otimes A$-null set.
\end{lemma}

\begin{lemma}\label{l:martingale}
$X$ is a uniformly integrable martingale if and only if it is a $\sigma
$-martingale of class (D).
\end{lemma}

\begin{lemma}\label{l:Pstarmartingale}
Let $P^\star\sim P$ be a probability measure with density process~$Z$.
A~real-valued semimartingale $X$ is a
\mbox{$P^\star$-$\sigma$-}martingale if and only if $XZ$ is
a~\mbox{$P$-$\sigma$-}martingale.
\end{lemma}

\begin{lemma}\label{l:girsanov}
Let $X$ be a $\rr^d$-valued semimartingale and let $P^\star\sim P$ be a
probability measure with density process $Z=Z_0\EEE(N)$. Denote by
\[
(b^{X,N},c^{X,N},F^{X,N},A) = \left(\pmatrix{b^X\cr b^N}, \pmatrix{
c^{X} & c^{XN} \cr   c^{NX} & c^{N}}, F^{X,N},A \right)
\]
differential characteristics of the $\rr^{d+1}$-valued seminartingale
$(X,N)$ relative to some truncation function $h$.
Then a version of the $P^\star$-differential characteristics of $(X,N)$
is given by
$(b^{X,N\star},c^{X,N\star},F^{X,N^\star},A)$,
where
\begin{eqnarray*}
b^{X,N\star}&=&b^{X,N}+c^{XN}+\int h(x,y)yF^{X,N}\bigl(d(x,y)\bigr),
\\
c^{X,N\star}&=&c^{X,N},
\\
{dF^{X,N\star}\over dF^{X,N}}(x,y)&=&1+y.
\end{eqnarray*}
\end{lemma}

\begin{lemma}
If $X$ is a $\sigma$-martingale and $\vartheta\in L(X)$, then
$\vartheta \mal X$ is a~\mbox{$\sigma$-}martingale as well.
\end{lemma}

\begin{lemma}\label{l:easily}
Let $X$ be a $\rr^d$-valued semimartingale and $\vartheta$ an
$\rr^d$-valued predictable process. Then $\vartheta\in L(X)$ if and
only if there exists a semimartingale $Z$ with $Z_0=0$ and an
increasing sequence $(D_n)_{n\in\nn}$ of predictable sets such that
$D_n\uparrow\Omega\times\rp$ up to an evanescent set,
$\vartheta1_{D_n}$ is bounded, and $1_{D_n}\mal
Z=(\vartheta1_{D_n})\mal X$ for any $n\in\nn$. In this case
$Z=\vartheta\mal X$.
\end{lemma}
\end{appendix}

\section*{\texorpdfstring{Acknowledgments.}{Acknowledgments}}
Parts of this research were done while the second author was visiting
Helsinki University of Technology. He wants to thank Esko Valkeila for
his hospitality. Both authors thank the Isaac Newton Institute for the
opportunity to work on the subject during the program on quantitative
finance. We are grateful to Thorsten Rheinl\"ander, Christophe
Stricker, Shige Peng, Jean Jacod, and Martin Schweizer for valuable
comments or discussions. Thanks are also due to an anonymous referee
for his detailed suggestions which enhanced the presentation of the
results.

\printaddresses

\end{document}